\documentclass[twocolumn,10pt]{asme2ej}
	
\usepackage[utf8]{inputenc}
\usepackage{textcomp}
\usepackage{url}
\usepackage{mathtools}
\usepackage{subfig}
\usepackage{floatrow}
\usepackage{amssymb}
\usepackage{stackengine}
\usepackage[outline]{contour}

\newcommand{\myref}[1]{(\ref{#1})}
\newcommand{\lnp}[1]{\log{\left(#1\right)}}
\newcommand{\myquote}[1]{``#1''}
\newcommand{\myspace}[1]{\hspace{#1em}}
\newcommand{\myunit}[2]{$#1\myspace{0.1}\mathrm{#2}$}
\newcommand{\mytilde}[0]{\raisebox{0.5pt}{\large\texttildelow}}

\newtheorem{definition}{Definition}[section]
\newtheorem{assumption}{Assumption}[section]
\newtheorem{prop}{Proposition A.\hspace{-0.25em}}
\newtheorem{remark}{Remark A.\hspace{-0.25em}}

\title{Mean-value exergy modeling of internal combustion engines: characterization\\ of feasible operating regions}

\author{Gabriele Pozzato
    \affiliation{
	Energy Resources Engineering\\
	Stanford University\\
	Stanford, CA 94305\\
    Email: gpozzato@stanford.edu
    }	
}

\author{Denise M. Rizzo
    \affiliation{ 
	Ground Vehicle Systems Center\\ U.S. Army CCDC\\
	6501 E. 11 Mile Road, Warren, MI 48397\\
	Email: denise.m.rizzo2.civ@mail.mil
    }
}

\author{Simona Onori 
    \affiliation{
	Energy Resources Engineering\\
	Stanford University\\
	Stanford, CA 94305\\
	Email: sonori@stanford.edu
    }
}

\begin{document}

\maketitle    

\begin{abstract}
{\it In this paper, a novel mean-value exergy-based modeling framework for internal combustion engines is developed.  The characterization of combustion irreversibilities,  thermal exchange between the in-cylinder mixture and the cylinder wall,  and non-stoichiometric combustion allows for a comprehensive description of the availability transfer and destruction phenomena in the engine. The model is applicable to internal combustion engines operating both in steady-state and over a sequence of operating points  and can be used to characterize the whole engine operating region, allowing to create static maps describing the exergetic behavior of the engine as a function of speed and load. The application of the proposed modeling strategy is shown for a turbocharged diesel engine. Ultimately, the static maps, while providing insightful information about inefficiencies over the whole operating field of the engine, are the enabling step for the development of exergy-based control strategies aiming at minimizing the overall operational losses of ground vehicles. }
\end{abstract}

\begin{nomenclature}
\entry{$\varepsilon$}{Second law efficiency $(\mathrm{-})$}
\entry{$\gamma$}{Specific heats ratio $(\mathrm{-})$}
\entry{$\lambda$}{Air-fuel equivalence ratio $(\mathrm{-})$}
\entry{$b$}{Hohenberg model coefficient $(\mathrm{-})$}
\entry{$c,d$}{Wiebe function coefficients $(\mathrm{-})$}
\entry{$CN$}{Cetane number $(\mathrm{-})$}
\entry{$f$}{Mole fraction $(\mathrm{-})$}
\entry{$N$}{Number of cycles $(\mathrm{-})$}
\entry{$n_{cyl}$}{Number of cylinders $(\mathrm{-})$}
\entry{$r_{c}$}{Compression ratio $(\mathrm{-})$}
\entry{$x,y$}{Fuel chemical formula coefficients $(\mathrm{-})$}
\entry{$x_{EGR}$}{EGR rate $(\mathrm{-})$}
\entry{$x_{fb}$}{Fuel burning rate (Wiebe function) $(\mathrm{-})$}
\entry{$t,dt$}{Time and its differential $(\mathrm{s})$}
\entry{$t_{cycle}$}{Engine cycle time $(\mathrm{s})$}
\entry{$\xi$}{Extent of reaction $(\mathrm{1/s})$}
\entry{$a$}{Crank radius $(\mathrm{m})$}
\entry{$l$}{Rod length $(\mathrm{m})$}
\entry{$s_{cp}$}{Distance between crank shaft and piston $(\mathrm{m})$}
\entry{$B$}{Bore $(\mathrm{m})$}
\entry{$L$}{Stroke $(\mathrm{m})$}
\entry{$A$}{Combustion chamber area $(\mathrm{m^2})$}
\entry{$A_p$}{Cylinder head surface area $(\mathrm{m^2})$}
\entry{$A_{ch}$}{Piston crown surface area $(\mathrm{m^2})$}
\entry{$V$}{Combustion chamber volume $(\mathrm{m^3})$}
\entry{$V_{d,tot}$}{Engine displacement $(\mathrm{m^3})$}
\entry{$S_p$}{Mean piston speed $(\mathrm{m/s})$}
\entry{$\mathrm{AFR}$}{Air-fuel ratio $(\mathrm{mol})$}
\entry{$\nu$}{Stoichiometric coefficient $(\mathrm{mol})$}
\entry{$\text{det}$}{Summation of stoichiometric coefficients $(\mathrm{mol})$}
\entry{$n,\dot{n}$}{Moles and molar flow rate  $(\mathrm{mol}),(\mathrm{mol/s})$}
\entry{$m,\dot{m}$}{Mass and mass flow rate $(\mathrm{kg}),(\mathrm{kg/s})$}
\entry{$M$}{Molar mass $(\mathrm{kg/mol})$}
\entry{$FMEP$}{Friction mean effective pressure $(\mathrm{Pa})$}
\entry{$C_1,C_2,C_3$}{FMEP coefficients $(\mathrm{kPa})$,$(\mathrm{s\ kPa})$,$(\mathrm{s^2kPa/m^2})$}
\entry{$P_0$}{Reference state pressure $(\mathrm{bar})$}
\entry{$P_I$}{Intake gas pressure $(\mathrm{bar})$}
\entry{$P_{cyl}$}{In-cylinder gas temperature $(\mathrm{bar})$}
\entry{$\mathcal{P}_{cyl}$}{In-cylinder pressure, crank-angle domain $(\mathrm{bar})$}
\entry{$T_0$}{Reference state temperature $(\mathrm{K})$}
\entry{$T_I,T_E$}{Intake and exhaust gas temperature $(\mathrm{K})$}
\entry{$T_w$}{Average wall temperature $(\mathrm{K})$}
\entry{$T_{cyl}$}{In-cylinder gas temperature $(\mathrm{K})$}
\entry{$\mathcal{T}_{cyl}$}{In-cylinder gas temperature, crank-angle domain $(\mathrm{K})$}
\entry{$Q_{cyl}$}{In-cylinder gas to wall thermal exchange  $(\mathrm{J})$}
\entry{$\mathcal{Q}_{cyl}$}{In-cylinder gas to wall thermal exchange, crank-angle domain $(\mathrm{J})$}
\entry{$\mathcal{Q}_{f}$}{Heat release during combustion $(\mathrm{J})$}
\entry{$\mu$}{Chemical potential $(\mathrm{J/mol})$}
\entry{$\psi_{ch},\psi_{ph}$}{Chemical and physical exergy flux $(\mathrm{J/mol})$}
\entry{$E_a$}{Apparent activation energy $(\mathrm{J/mol})$}
\entry{$\mathrm{ex}^{ch}$}{Specific chemical exergy $(\mathrm{J/mol})$}
\entry{$g$}{Gibbs free energy $(\mathrm{J/mol})$}
\entry{$h$}{Specific enthalpy $(\mathrm{J/mol})$}
\entry{$R_{gas}$}{Ideal gas constant $(\mathrm{J/(mol\ K)})$}
\entry{$s$}{Specific entropy $(\mathrm{J/(mol\ K)})$}
\entry{$a_f$}{Fuel specific chemical exergy $(\mathrm{MJ/kg})$}
\entry{$LHV$}{Fuel lower heating value $(\mathrm{MJ/kg})$}
\entry{$\mathcal{X},\dot{\mathcal{X}}$}{Exergy in the crank-angle domain $(\mathrm{J}),(\mathrm{W})$}
\entry{$X,\dot{X}$}{Exergy and exergy rate $(\mathrm{J}),(\mathrm{W})$}
\entry{$S,\dot{S}$}{Entropy and entropy rate $(\mathrm{J/K}),(\mathrm{W/K})$}
\entry{$P_{eng}$}{ICE power $(\mathrm{W})$}
\entry{$h_{cyl}$}{Convective heat transfer coefficient $(\mathrm{kW/(m^2K)})$}
\entry{$\theta,d\theta$}{Crank-angle and its differential $(\mathrm{rad})$}
\entry{$\theta_{corr}$}{Crank-angle correction $(\mathrm{rad})$}
\entry{$\theta_{EC}$}{Crank-angle at EC $(\mathrm{rad})$}
\entry{$\theta_{SC}$}{Crank-angle at SC $(\mathrm{rad})$}
\entry{$\theta_{SI}$}{Crank-angle at SI $(\mathrm{rad})$}
\entry{$\theta_{TDC}$}{Crank-angle at TDC $(\mathrm{rad})$}
\entry{$\Delta\theta$}{Combustion duration $(\mathrm{rad})$}
\entry{$\tau_{id}$}{Ignition delay $(\mathrm{rad})$}
\entry{$\omega_{eng}$}{ICE rotational speed $(\mathrm{rad/s})$}
\entry{$\mathrm{T}_{eng}$}{ICE $(\mathrm{Nm})$}

\section*{Notation}
\entry{$0$}{Reference state}
\entry{$\star$}{Restricted state}
\entry{$\sigma$}{Chemical species}
\entry{$\iota$}{Index indicating intake and exhaust: $\iota\in\{I,E\}$}
\entry{$j$}{Engine operating point}
\entry{$k$}{Index to indicate an exergy term in a set $\mathcal{K}$}
\entry{$\mathcal{K}$}{Set collecting exergy terms}
\entry{$\mathcal{S}$}{Set collecting the chemical species $\sigma$, defined as $\{N_2,CO_2,H_2O,O_2\}$}
\entry{$d\chi/d\theta$}{Crank-angle derivative of a variable $\chi$}
\entry{$\dot{\chi}$}{Time derivative of a variable $\chi$}
\entry{$a,f,s$}{Subscripts for air, fuel, and stoichiometric, respectively}
\entry{$exh$}{Exhaust gas introduced in the engine by the EGR}
\entry{I,E}{Intake and exhaust}
\entry{EGR}{Exhaust gas recirculation}
\entry{EC}{End of combustion}
\entry{SC}{Start of combustion}
\entry{SI}{Start of injection}
\entry{BDC}{Bottom dead center}
\entry{TDC}{Top dead center}
\end{nomenclature}

\section{Introduction}
In the quest for sustainable solutions, increasing fuel economy, reducing vehicular emissions, and improving energy efficiency are mandatory actions.  In this context, military ground vehicles are migrating from stand alone internal combustion engine (ICE) powertrains to electrified ones \cite{tadjdeh2019army,mamun2018integrated},  to improve the overall energy efficiency while reducing operational costs and noise emissions for stealthier operations. To unlock the potential of these technologies, formal modeling and analysis techniques must be developed for a complete understanding of the irreversibilities degrading the efficiency of the system \cite{james2020review}. 

In this framework, availability, or exergy, is a useful metric to quantify the thermal, mechanical, and chemical work a system can perform with respect to a reference state, usually called environment.  Differently from approaches based on the first law of thermodynamics,  rooted in the principle of conservation of energy,  exergy-based modeling implements the second law of thermodynamics exploiting the concept of entropy. This allows for the explicit quantification of the irreversibilities of the system and, ultimately, for efficiency improvement.

Exergy-based modeling has been used for system design and optimization in many engineering fields.  For example,  in aerospace engineering, the concept of exergy has been widely used to produce more effective designs, integrating and balancing diverse systems. In \cite{gilbert2016uses} and \cite{watson2018system}, the authors propose an availability analysis for rockets and launch vehicles, respectively. The design and optimization of hypersonic aircrafts is tackled in \cite{camberos2009systems} and \cite{riggins2006methodology}. Exergy analysis is applied also in other fields, such as naval engineering \cite{baldi2018energy} and power generation \cite{rosen1986energy,rosen2001energy}. Moreover, availability concepts are used for the analysis of combustion processes in engines. ICEs are complex systems composed of up to 2000 interacting parts subjected to friction, heat transfer, and thermo-mechanical stresses. Controlling in-cylinder combustion phenomena is imperative to optimize braking work generation, while minimizing inefficiencies. Considering spark ignition engines, \cite{rakopoulos2008availability} and \cite{valencia2019energy} analyze the exergy transfer and destruction processes in engines fueled with synthetic and natural gas, respectively. For what concerns compression ignition engines, an overview on exergy modeling for both naturally aspirated and turbocharged diesel engines is provided in \cite{rakopoulosexergyengine}. In \cite{alkidas1988application,van1990second} and \cite{giakoumis2007irreversibility}, steady-state and transient operating conditions are analyzed, respectively. The authors of \cite{canakci2006energy} examine the availability balance of a diesel engine fueled with different biodiesels. In \cite{sayin2016energy}, a similar analysis is performed considering biodiesel, diesel, and bioethanol blends. 

On the other hand,  some works show the application of exergy principles for the development of control algorithms aiming at maximizing the efficiency of the system.  In \cite{razmaraexergyhvac}, \cite{trinklein2020modeling}, and \cite{trinklein2020exergy}, the effectiveness of exergy-based control algorithms is shown for air conditioning in buildings, ships, and aircrafts, respectively. In the context of ground vehicles' powertrain,  \cite{razmaraexergyengine} successfully optimizes the in-cylinder combustion in homogeneous charge compression ignition (HCCI) engines relying on model predictive control (MPC),  achieving 6.7\% fuel saving. 

The scarcity of exergy-based analysis for ground vehicles has motivated the authors of this paper to develop a comprehensive exergy-based modeling framework for hybrid and electric vehicles (HEVs and EVs),  see \cite{ourpaper}.  This framework allows for the quantification of exergy transfer and destruction phenomena within the energy storage and conversion devices of the powertrain.  For the exergetic characterization of HEVs,  this framework relies on a simplified ICE model.  In particular, the thermal exchange between the in-cylinder mixture and the cylinder wall is modeled relying on the Taylor\&Toong correlation. In accordance with \cite{ parra2008heat},  this correlation can be used for an average description of the engine heat transfer but suffers from two drawbacks.  On the one hand,  for its computation the in-cylinder mixture temperature, viscosity, and conductivity as a function of the air-fuel ratio are needed (quantities difficult to measure/estimate and usually retrieved from \cite{taylor1985internal}).  On the other hand,  this correlation can not provide detailed insights on the engine thermal behavior because formulated in the time rather than crank-angle domain.  Moreover,  the ICE model in \cite{ourpaper} does not explicitly consider the exergy destruction related to the combustion process, with combustion irreversibilities lumped together with other terms, namely,  blow-by gases,  unburnt fuel,  and intake air flow.  Also,  this ICE model is developed under the assumption of stoichiometric combustion -- in other words,  it assumes that the air entering the cylinder burns all fuel and no excess air is left -- and is valid for spark ignition engines only.  Ultimately, the engine model proposed in \cite{ourpaper},  while being suitable for the description of the exergetic behavior at the powertrain-level,  poses some limitations for what concerns the quantification of the ICE exergy transfer and destruction phenomena, and lacks in the modeling of combustion irreversibilities (the principal source of exergy destruction).  This, together with the fact that ICEs are the principal source of inefficiency, makes imperative the development of a more sophisticated and accurate ICE exergy model.

In this paper,  we extend the engine exergy-based model proposed in \cite{ourpaper}, formulating a thorough model for exergy quantification in ICEs.  For the first time, a detailed methodology for the derivation of a mean-value\footnote{A mean-value model does not consider the engine reciprocating behavior and assumes all the thermodynamical phenomena to happen continuously. } exergy-based model for ICEs, accounting for all the availability transfer and destruction phenomena, is proposed. In particular, a characterization of combustion irreversibilities, based on the analysis of the chemical potentials, is provided.  Moreover,  the thermal exchange between the in-cylinder mixture and the cylinder wall is modeled relying on the Hohenberg correlation, which allows for a more detailed characterization of the heat transfer during combustion (if compared to the Taylor\&Toong correlation),  and combustion is analyzed also for non-stoichiometric operation.  The ICE exergy model is extended to characterize the overall engine operating region in the speed-load plane, allowing to derive maps describing the quasi-static exergetic behavior of the engine as a function of speed and load. These maps, computed once for the engine under study, can be used for the development of \myquote{exergy management strategies} as they provide useful information on the engine efficiency (and inefficiency) for each operating condition while keeping the computational burden low,  an important property in particular for real-time applications.  The approach is here developed starting from a turbocharged compression ignition diesel engine, equipped with exhaust gas recirculation (EGR).  However, the procedure is general and can be used in any device equipped with an ICE (not only internal combustion engine vehicles (ICEVs) and HEVs,  but also ships,  light aircrafts,  power units,  etc.).  In particular,  the same methodology can be applied to any compression ignition engine, regardless of the fueling used, and spark ignition engines. In the latter scenario, the fuel is injected during the intake stroke and the combustion is triggered by the spark plug, without any ignition delay. 

The remainder of the paper is organized as follows. In Section \ref{sec:definitions}, fundamental concepts related to availability modeling are summarized.  In Section \ref{sec:ice_model}, the ICE model, in terms of airflow path, combustion reaction, and in-cylinder dynamics, is presented. Moreover, the derivation of average in-cylinder pressure, heat transfer, and temperature maps is presented. Section \ref{sec:exergy_model} formalizes the exergy-based modeling of the ICE, introducing all the sources of exergy transfer and destruction phenomena. In Section \ref{sec:results},  results for the engine operating at steady-state and over a sequence of operating points are shown, and static maps describing the exergetic behavior of the engine are derived.  Finally, conclusions are carried out in Section \ref{sec:conclusions}.

\section{Exergy modeling: definitions}\label{sec:definitions}
In this section, key definitions related to exergy modeling are introduced.  For a comprehensive summary of exergy concepts,  readers are referred to \cite{ourpaper}.

\begin{definition}[Exergy]\label{def:EX}
Exergy (or availability) is the maximum useful work that can be extracted from a system at a given state, with respect to a thermodynamic and chemical reference state. 
\end{definition}

\begin{definition}[Physical exergy]\label{def:EXph}
Work potential between the current and restricted state of the system.
\end{definition}

\begin{definition}[Chemical exergy]\label{def:EXch}
Work potential accounting for the different chemical composition between the restricted and reference state.
\end{definition}

\begin{definition}[Reference state]\label{def:DS}
The reference state (also referred as dead state or environment) is characterized by a pressure $P_0$, a temperature $T_0$, and a mixture of chemical species combined according to the molar fractions $f_0$. At the reference state, the entropy of the system is maximized and the available work (chemical, thermodynamical, mechanical, etc.) is zero.
\end{definition}

\begin{definition}[Restricted state]\label{def:RS}
The restricted state indicates a system not in chemical equilibrium with the reference state.
\end{definition}

From now on, quantities expressed with respect to the reference and restricted state are denoted by subscript $0$ and superscript $\star$, respectively.

\section{Internal combustion engine modeling}\label{sec:ice_model}
In this section, the mean-value model for the ICE is presented.  The derivation of the model is carried out under the following fundamental assumptions.
 \begin{assumption} The gaseous mixtures are composed by ideal gases only.\end{assumption} 
 \begin{assumption}\label{assum:fuelburn} The combustion process burns all the available fuel, thus, no unburnt fuel is present in the exhaust gases. \end{assumption} 
 \begin{assumption} \label{assum:ss} Engine operating points are analyzed in steady-state, thus, exhaust transport and torque actuation delays are not modeled.\end{assumption} 
 \begin{assumption} \label{assum:ss_incyl} At a given operating point, the cylinders composing the engine behave in exactly the same manner.\end{assumption} 
  \begin{itemize}
 \item[$\square$] This assumption is a direct consequence of the mean-value modeling approach,  where the reciprocating behavior of the whole engine is averaged over time.
 \end{itemize}
 
 \subsection{Engine technical specifications}\label{sec:icespecs}
\begin{figure}[!t]
\centering
\includegraphics[width = \columnwidth]{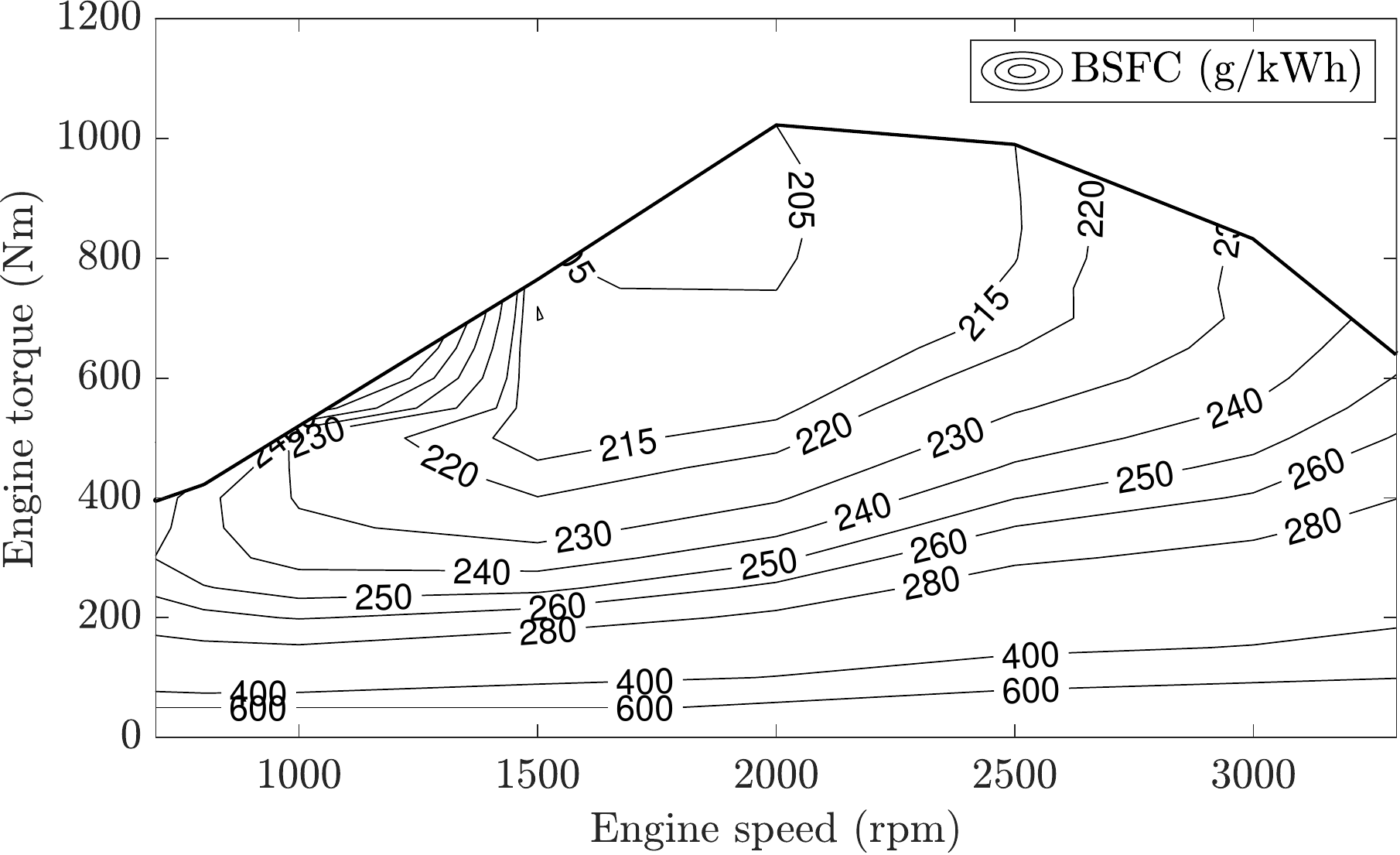}
\caption{BSFC map \cite{liu2018combined}.}
\label{fig:bsfc}
\end{figure}

A Power Stroke 6.4L V8 diesel engine is considered.  The engine is turbocharged, equipped with an EGR system, and characterized by a peak power of \myunit{260}{kW} at \myunit{3000}{rpm} \cite{engineprop}.  A comprehensive list of the ICE parameters is shown Table \ref{tab:param1}.  Moreover, the brake-specific fuel consumption (BSFC) map,  from \cite{liu2018combined}, is shown in Figure \ref{fig:bsfc}. 

\begin{table}[t!]
	\centering
	\begin{center}
		\caption{ICE parameters.}
		\label{tab:param1}
		\resizebox{0.9\columnwidth}{!}{
		\begin{tabular}{cccc}
			\textbf{Parameter} & \textbf{Unit} & \textbf{Value} & \textbf{Reference} \\ \hline
			$b$ & (-) & 1.4 &  \scriptsize{\cite{lakshminarayanan2010modelling}}\\
			$c$ & (-) &  $-\log(0.001)$ & \scriptsize{\cite{guzzellabook}} \\
			$d$ & (-) & 2 & \scriptsize{\cite{widd2012single}} \\
		    $f_{N_2,0}$ & (-) & 0.7567 & \scriptsize{\cite{razmaraexergyengine}} \\
			$f_{CO_2,0}$ & (-) & 0.0003 & \scriptsize{\cite{razmaraexergyengine}}  \\
			$f_{H_2O,0}$ & (-) & 0.0303 & \scriptsize{\cite{razmaraexergyengine}} \\
			$f_{O_2,0}$ & (-) & 0.2035 &\scriptsize{\cite{razmaraexergyengine}}  \\
			$f_{others,0}$ & (-) & 0.0092 & \scriptsize{\cite{razmaraexergyengine}} \\
			$r_c$ & (-) & 17.5:1 & \scriptsize{\cite{engineprop}} \\
			$n_{cyl}$ & (-) & 8 & \scriptsize{\cite{engineprop}} \\
			$CN$ & (-) & 50 & \scriptsize{\cite{heywoodbook}}\\
			$x_{EGR}$ & (-) & 0.2 (or 20\%) & \scriptsize{\cite{rakopoulos2009diesel}}\\
			$x$  & (-) & 14.4 & \scriptsize{\cite{rakopoulosexergyengine}} \\
			$y$ & (-) &  24.9 & \scriptsize{\cite{rakopoulosexergyengine}} \\
			$a$ & (m) & 52.5$\times 10^{-3}$ & \scriptsize{\cite{engineprop}}\\
			$l$ & (m) &  210$\times 10^{-3}$ & \scriptsize{\cite{engineprop}}\\
			$B$ & (m) & 98.2$\times 10^{-3}$ & \scriptsize{\cite{engineprop}}\\
			$L$ & (m) & 105$\times 10^{-3}$ & \scriptsize{\cite{engineprop}}\\
		     $V_{d,tot}$ & ($\mathrm{l}$) & 6.4 & \scriptsize{\cite{engineprop}}\\
			$T_0$ & (K) & 298.15 & -\\
			$T_I$ & (K) & 323.15 & -\\
			$T_w$ & (K) & 400 & \scriptsize{\cite{vzak2016cylinder}} \\
			$P_0$ & (bar) & 1 & -\\
			$P_I$ & (bar) & 1 & -\\
			$C_1$ & ($\mathrm{kPa}$) & 75 & \scriptsize{\cite{heywoodbook}}\\
			$C_2$ & ($\mathrm{s\ kPa}$) & 0.458 & \scriptsize{\cite{heywoodbook}} \\                
			$C_3$ & ($\mathrm{s^2kPa/m^2}$) & 0.4 & \scriptsize{\cite{heywoodbook}} \\
			$\theta_{SI}$ & (rad) & 15.7$\times (\pi/180)$ & \scriptsize{\cite{widd2012single}} \\
			& & before TDC &  \\
			$\theta_{TDC}$ & (rad) & 0 & -\\
			$\Delta\theta$ & (rad) & 70$\times (\pi/180)$ & \scriptsize{\cite{punov2017study}}\\
		    $M_{f}$ & (kg/mol) & 0.198 & - \\
			$M_{N_2}$ & (kg/mol) & 0.028 & - \\
			$M_{CO_2}$ & (kg/mol) & 0.044 & - \\
			$M_{H_2O}$ & (kg/mol) & 0.018 & - \\
			$M_{O_2}$ & (kg/mol) & 0.032 & - \\
		     $\mathrm{ex}^{ch}_{CO_2}$ & (J/mol)  & 19870 &\scriptsize{\cite{bejan1999thermodynamic}}\\ 
		     $\mathrm{ex}^{ch}_{H_2O}$ & (J/mol)  & 900 & \scriptsize{\cite{bejan1999thermodynamic}}\\ 
		     $\mathrm{ex}^{ch}_{O_2}$ & (J/mol)  & 3970 & \scriptsize{\cite{bejan1999thermodynamic}}\\ 
			$R_{gas}$ & ($\mathrm{J/(mol\ K)}$) & 8.31 & - \\
			$LHV$ & ($\mathrm{MJ/kg}$) & 42.50 & \scriptsize{\cite{guzzellabook}} \\
			\hline 
		\end{tabular}}
	\end{center}
\end{table}

\subsection{Airflow path}
\begin{figure}[!tb]
\centering 
\subfloat[Fuel rate]{\includegraphics[width=1\columnwidth]{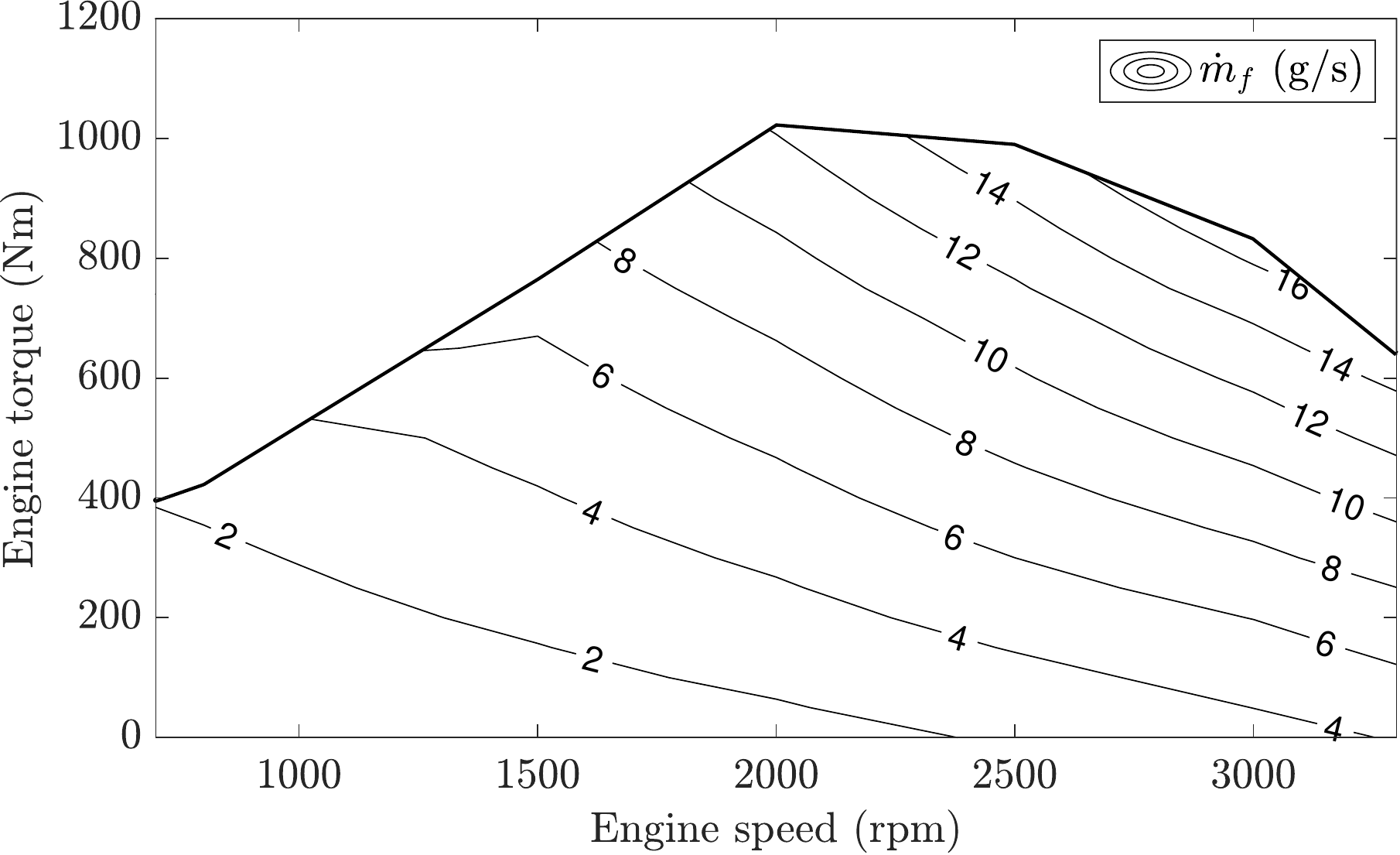}}\\
\subfloat[Air-fuel equivalence ratio]{\includegraphics[width=1\columnwidth]{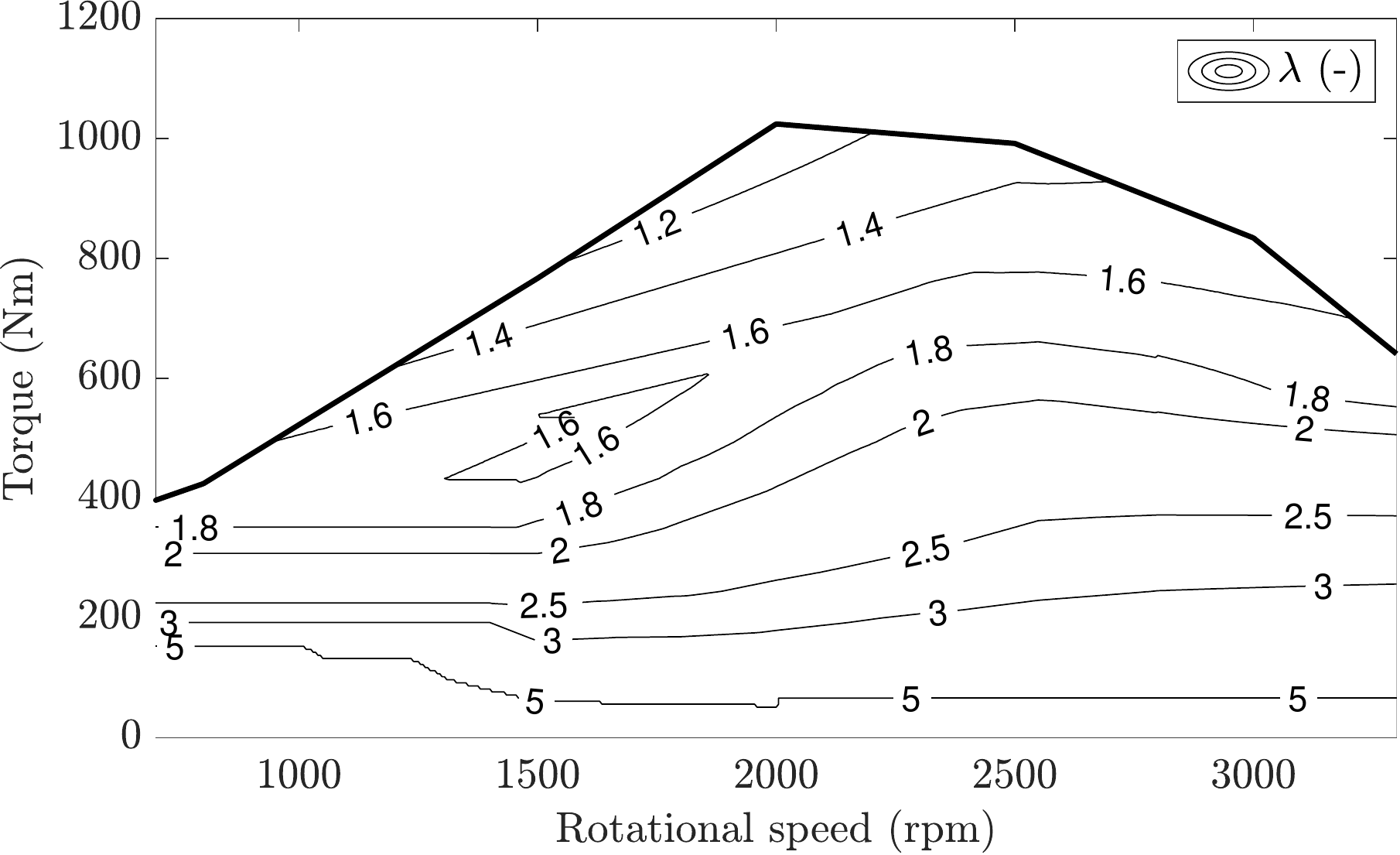}}\\
\subfloat[Exhaust gas temperature]{\includegraphics[width=1\columnwidth]{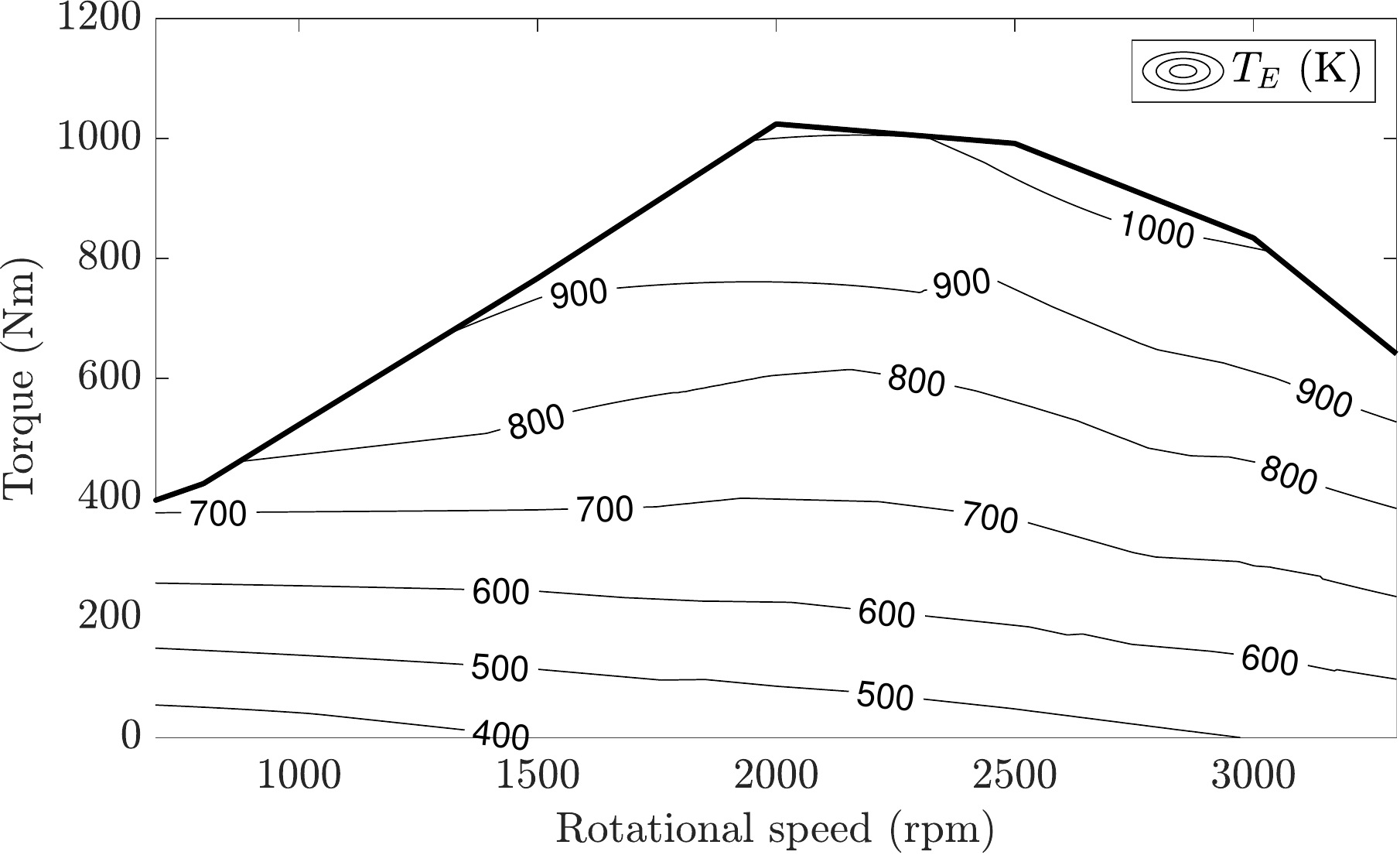}}
\caption{Fuel rate ($\dot{m}_f$), air-fuel equivalence ratio ($\lambda$), and exhaust gas temperature ($T_E$) maps, for the diesel engine in Section \ref{sec:icespecs},  as a function of speed and torque.  As shown in map (b),  $\lambda$ is always greater than 1,  reasonable for a diesel engine.}
\label{fig:map_lambda_fuel}	
\end{figure}

The airflow path is modeled considering the mass flow rates at the intake and exhaust manifolds, under the presence of EGR.  The air-fuel ratio is defined according to the following equation:
\begin{equation}
\text{AFR}(t) = \frac{\dot{m}_a(t)}{\dot{m}_f(t)}
\label{eq:AFR}
\end{equation}
with $\dot{m}_a$ and $\dot{m}_f$ the intake airflow and fuel rates, respectively.  Starting from Equation \myref{eq:AFR}, the air-fuel equivalence ratio is introduced:
\begin{equation}
\lambda(t) = \frac{\text{AFR}(t)}{\text{AFR}_s}
\label{eq:lambda}
\end{equation}
where $\mathrm{AFR}_s$ is the stoichiometric air-fuel ratio,  i.e.,  the ideal ratio of air that burns all fuel with no excess air.  The air-fuel equivalence ratio and the fuel rate are functions of the engine operating point:
\begin{equation}
\begin{split}
&\dot{m}_f(t) = \mathrm{f}_{f}(\omega_{eng}(t),\mathrm{T}_{eng}(t))\\
&\lambda(t) = \mathrm{f}_{\lambda}(\omega_{eng}(t),\mathrm{T}_{eng}(t))
\end{split}
\label{eq:maps_lamfr}
\end{equation}
with $\mathrm{T}_{eng}$ and $\omega_{eng}$ the engine torque and rotational speed, respectively. In this work, $\dot{m}_f$ and $\lambda$ are obtained from maps (a) and (b) in Figure \ref{fig:map_lambda_fuel}. For a diesel engine, the air-fuel equivalence ratio $\lambda$ is always greater than 1, i.e., the combustion takes place in lean conditions. From Equation \myref{eq:lambda} one can compute $\dot{m}_a$,  therefore the intake flow rate is derived as:
\begin{equation}
\dot{m}_I(t)=\dot{m}_a(t)+\dot{m}_{exh}(t)
\label{eq:intk_gas}
\end{equation}
where $\dot{m}_{exh}$ is the portion of exhaust gas introduced by the EGR\footnote{The exhaust gas is assumed to be composed by $CO_2$ and $H_2O$; $NO_x$ and $CO$ are neglected given their low concentration with respect to the other species \cite{guzzellabook}.}. According to \cite{guzzellabook}, $\lambda$ is defined after the perfect mixing between the recirculated gases and fresh air, thus, it already accounts for the presence of recirculated oxygen and nitrogen. Recalling that $\dot{m}_{exh}(t) = x_{EGR}\ \dot{m}_I(t)$ and $\dot{m}_{a}(t) = (1-x_{EGR})\dot{m}_I(t)$ \cite{guzzellabook}, Equation \myref{eq:intk_gas} is rewritten as follows:
\begin{equation}
\begin{split}
\dot{m}_I(t)&=\dot{m}_a(t)\left(1+\frac{x_{EGR}}{1-x_{EGR}}\right)=\\&=\dot{m}_a(t)\frac{1}{1-x_{EGR}}
\end{split}
\label{eq:intk_kgs}
\end{equation}
where $x_{EGR}$ is the EGR rate, i.e., the portion of recirculated gas. Finally, the exhaust manifold flow rate is obtained as:
\begin{equation}
\dot{m}_{E}(t) = \dot{m}_{I}(t) + \dot{m}_{f}(t)
\label{eq:exh_kgs}
\end{equation}
The intake mixture enters the engine at a constant temperature $T_I =$ \myunit{323.15}{K} (a reasonable average value for turbocharged diesel engines \cite{di2018optimization}), instead, the exhaust gas temperature $T_E$ varies according to the engine operating condition (Figure \ref{fig:map_lambda_fuel}c). As a reference,  maps similar to those depicted in Figure \ref{fig:map_lambda_fuel} (for both compression and spark ignition engines) can be found in the MathWorks Powertrain Blockset toolbox \cite{powertrainblockset}.

\subsection{Combustion reaction}
According to \cite{asad2014exhaust}, the combustion reaction in the presence of EGR takes the following form:
\begin{equation}
\resizebox{1\columnwidth}{!}{$
\begin{split}
&C_xH_y+\frac{\lambda(t)}{1-x_{EGR}}\left(x+\frac{y}{4}\right)3.76N_2+\frac{x_{EGR}}{1-x_{EGR}}\left(xCO_2+\frac{y}{2}H_2O\right)+\\
&+\left(x+\frac{y}{4}\right)\frac{\lambda(t)-x_{EGR}}{1-x_{EGR}}O_2\ \contour{black}{$\rightarrow$}\ \frac{1}{1-x_{EGR}}\bigg[xCO_2+\frac{y}{2}H_2O+\\
&+3.76\lambda(t) \left(x+\frac{y}{4}\right)N_2+(\lambda(t)-1)\left(x+\frac{y}{4}\right)O_2\bigg]
\end{split}$}
\label{eq:comb_react}
\end{equation}
where the coefficients $x$ and $y$ define the fuel ($C_xH_y$) composition. According to \cite{rakopoulosexergyengine}, $x = 14.4$ and $y=24.9$ describe the average composition of diesel with good accuracy. Given a certain operating condition, the reaction \myref{eq:comb_react} can be used to compute the mole fraction of the different species composing the intake and exhaust gas, respectively. Thus, the following variables are introduced:
\begin{equation}
\begin{split}
\text{det}_{I}(t) =&\  \frac{\lambda(t)}{1-x_{EGR}}\left(x+\frac{y}{4}\right)3.76+\frac{x_{EGR}}{1-x_{EGR}}\left(x+\frac{y}{2}\right)+\\
&+\left(x+\frac{y}{4}\right)\frac{\lambda(t)-x_{EGR}}{1-x_{EGR}}\\
\text{det}_{E}(t) =&\ \frac{1}{1-x_{EGR}}\bigg[x+\frac{y}{2}+3.76\lambda(t) \left(x+\frac{y}{4}\right)+\\
&+(\lambda(t)-1)\left(x+\frac{y}{4}\right)\bigg]
\end{split}
\end{equation}
with $\text{det}_{I}$ and $\text{det}_{E}$ the denominators used to compute the volume fractions for reactants (intake gases) and products (exhaust gases):
\begin{equation}
\begin{split}
&Intake\ gases\\
& \hspace{1em}f_{N_2}^I(t)= \frac{\frac{\lambda(t)}{1-x_{EGR}}\left(x+\frac{y}{4}\right)3.76}{\text{det}_I(t)}=\frac{\nu_{N_2}^I(t)}{\text{det}_I(t)},\\
& \hspace{1em}f_{CO_2}^I(t)= \frac{\frac{x_{EGR}}{1-x_{EGR}}x}{\text{det}_I(t)}=\frac{\nu_{CO_2}^I}{\text{det}_I(t)},\\
& \hspace{1em}f_{H_2O}^I(t) = \frac{\frac{x_{EGR}}{1-x_{EGR}}\frac{y}{2}}{\text{det}_I(t)} =\frac{\nu_{H_2O}^I}{\text{det}_I(t)},\\
& \hspace{1em}f_{O_2}^I(t) = \frac{\frac{\lambda(t)-x_{EGR}}{1-x_{EGR}}\left(x+\frac{y}{4}\right)}{\text{det}_I(t)}=\frac{\nu_{O_2}^I(t)}{\text{det}_I(t)}\\
\end{split}
\label{eq:gas_mix_I}
\end{equation}
\begin{equation}
\begin{split}
&Exhaust\ gases\\
& \hspace{1em}f_{N_2}^E(t) = \frac{\frac{\lambda(t)}{1-x_{EGR}}\left(x+\frac{y}{4}\right)3.76}{\text{det}_E(t)} = \frac{\nu_{N_2}^E(t)}{\text{det}_E(t)},\\
& \hspace{1em}f_{CO_2}^E(t) = \frac{\frac{1}{1-x_{EGR}}x}{\text{det}_E(t)}=\frac{\nu_{CO_2}^E}{\text{det}_E(t)},\\
& \hspace{1em}f_{H_2O}^E(t)= \frac{\frac{1}{1-x_{EGR}}\frac{y}{2}}{\text{det}_E(t)}=\frac{\nu_{H_2O}^E}{\text{det}_E(t)},\\
& \hspace{1em}f_{O_2}^E(t) = \frac{\frac{\lambda(t)-1}{1-x_{EGR}}\left(x+\frac{y}{4}\right)}{\text{det}_E(t)}=\frac{\nu_{O_2}^E(t)}{\text{det}_E(t)}\\
\end{split}
\label{eq:gas_mix_E}
\end{equation}
From Equations \myref{eq:gas_mix_I} and \myref{eq:gas_mix_E}, the molar masses of the gaseous mixtures are computed as follows:
\begin{equation}
\begin{split}
&M_I(t)= \sum_{\sigma\in\mathcal{S}} f_\sigma^I(t)M_\sigma,\quad M_E(t)= \sum_{\sigma\in\mathcal{S}} f_\sigma^E(t)M_\sigma,\\
&\sigma\in\mathcal{S}=\{N_2,CO_2,H_2O,O_2\}
\end{split}
\end{equation}
Starting from Equations \myref{eq:intk_kgs}, \myref{eq:exh_kgs}, \myref{eq:gas_mix_I}, and \myref{eq:gas_mix_E}, the molar flow rates for intake ($\dot{n}_{I}$) and exhaust ($\dot{n}_{E}$) manifolds are obtained:
\begin{equation}
\dot{n}_I(t) = \frac{\dot{m}_I(t)}{M_I},\quad \dot{n}_E(t)=\frac{\dot{m}_E(t)}{M_E}
\label{eq:mole_rate}
\end{equation}

\subsection{In-cylinder dynamics}\label{sec:engine_dyn_crank}
To obtain an effective mean-value description of the in-cylinder combustion phenomena,  we first carry out the modeling in the crank-angle domain. To this aim, a single-zone modeling approach, in which the working fluids form \textit{one} thermodynamic system undergoing energy and mass exchange with the surroundings, is employed. In this regard, the heat released by combustion is assumed to be evenly distributed throughout the cylinder and no distinction between the burnt/unburnt fraction of the mixture is made. Moreover, spatial homogeneity of pressure and temperature is assumed. According to different authors \cite{rakopoulos2009diesel,heywoodbook}, this is an effective approach to analyze the thermodynamic behavior of ICEs and provides a detailed description of pressure, heat transfer, and temperature. 

\subsubsection{Crank-angle resolved model}\label{sec:engine_dyn_crank_2}
\begin{figure}[!t]
\centering
\includegraphics[width = 0.7\columnwidth]{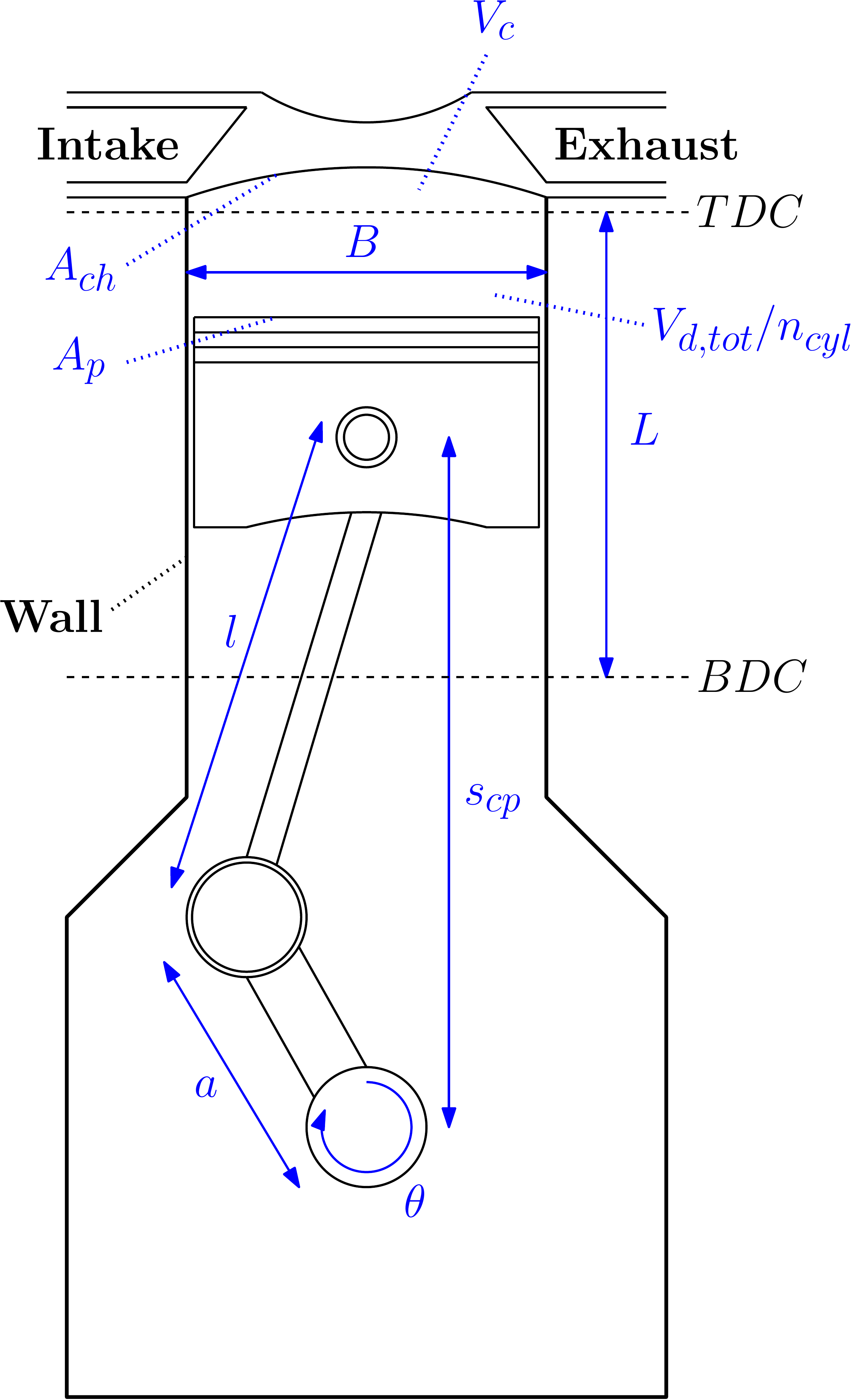}
\caption{Geometry of the cylinder,  piston,  connecting rod,  and crankshaft.  Parameters $s_{cp}$, $a$, $l$, $B$, and $L$ are the distance between crank shaft and piston, crank radius, rod length,  cylinder bore,  and stroke, respectively.  The clearance ($V_c$) and displacement volumes ($V_{d,tot}/n_{cyl}$),  defined with respect to the BDC and TDC,  are shown.  $A_{ch}$ and $A_p$ are the cylinder head and piston crown surface areas.}
\label{fig:enginecyldyn}
\end{figure}

To describe the in-cylinder phenomena, it is key to catch the engine reciprocating behavior.  The crank-angle variable $\theta$ is introduced and used to describe the motion of the piston inside the cylinder (depicted in Figure \ref{fig:enginecyldyn}) in terms of distance between the crank shaft and piston ($s_{cp}$), swept cylinder volume ($V$), and combustion chamber surface area ($A$):
\begin{equation}
\begin{split}
& s_{cp}(\theta) = a\ \cos\theta + \sqrt{l^2 -a^2\sin^2\theta}\\
& V(\theta) = V_c + \frac{\pi B^2}{4}(l+a-s_{cp}(\theta))=\\
& \hspace{2.15em}=\frac{V_{d,tot}/n_{cyl}}{r_c-1}+\frac{\pi B^2}{4}(l+a-s_{cp}(\theta))\\
& A(\theta ) = A_{ch} + A_p + \pi B(l+a-s_{cp}(\theta))=\\
& \hspace{2.1em}=2\left(\frac{\pi B^2}{4}\right)+\pi B(l+a-s_{cp}(\theta))
\end{split}
\label{eq:piston_motion}
\end{equation}
with $a$ the crank radius, $l$ the rod length, $B$ the cylinder bore, $n_{cyl}$ the number of cylinders, $V_{d,tot}$ the engine displacement, $V_c$ the clearance volume -- function of the cylinder displacement $V_{d,tot}/n_{cyl}$ and of the compression ratio $r_c$ --, and $A_{ch}$ and $A_p$ the cylinder head and piston crown surface areas, both equal to $(\pi B^2)/4$.

Starting from Equation \myref{eq:piston_motion}, the crank-angle derivatives of the swept cylinder volume and combustion chamber surface area are given by:
\begin{equation}
\begin{split}
& \frac{dV}{d\theta} = \frac{\pi B^2}{4}\left[ a\sin\theta\left( \frac{f\cos\theta}{\sqrt{1-f^2\sin^2\theta}} \right) \right]\\
& \frac{dA}{d\theta} =\pi B\left[ a\sin\theta\left( \frac{f\cos\theta}{\sqrt{1-f^2\sin^2\theta}} \right) \right]\\
\end{split}
\end{equation}
where $f=a/l$. Thus, relying on the first principle of thermodynamics, the following expression for the in-cylinder pressure $\mathcal{P}_{cyl}$ is obtained:
\begin{equation}
\frac{d\mathcal{P}_{cyl}}{d\theta}=\frac{\gamma-1}{V(\theta)}\left(\frac{d\mathcal{Q}_f}{d\theta}-\frac{d\mathcal{Q}_{cyl}}{d\theta}\right)-\frac{\gamma}{V(\theta)}\mathcal{P}_{cyl}(\theta)\frac{dV}{d\theta}
\label{eq:pressure_incyl}
\end{equation}
with $\gamma$ the specific heat ratio (equal to $1.4-0.16/\lambda$ for a diesel engine \cite{sanli2008influence}), $\mathcal{Q}_{cyl}$ the in-cylinder gas to wall heat exchange, and $\mathcal{Q}_{f}$ the heat released in the combustion process, defined as follows:
\begin{equation}
\frac{d\mathcal{Q}_f}{d\theta} = \frac{dx_{fb}}{d\theta}m_f LHV, \quad m_f = \dot{m}_f\frac{4\pi}{\omega_{eng}}\frac{1}{n_{cyl}}
\end{equation}
with $LHV$ the fuel lower heating value, $m_f$ the injected fuel per cylinder, and $\dot{m}_f$ the fuel rate computed from Equation \myref{eq:maps_lamfr} at the operating point $(\omega_{eng},\mathrm{T}_{eng})$, constant over one cycle. The fuel burning rate is computed from the Wiebe function \cite{rakopoulos2009diesel}:
\begin{equation}
\frac{dx_{fb}}{d\theta} = \frac{c(d+1)}{\Delta\theta}\left(\frac{\theta - \theta_{SC}}{\Delta\theta}\right)^d\exp\left[-c\left(\frac{\theta-\theta_{SC}}{\Delta\theta}\right)^{d+1}\right]
\end{equation}
where $\Delta\theta$ is the combustion duration, $c$ the combustion efficiency coefficient (controlling if the combustion process leaves unburnt fuel), and $d$ a shaping parameter. The start of combustion angle is defined as $\theta_{SC}=\theta_{SI}-\tau_{id}$, where $\theta_{SI}$ is the start of injection angle and $\tau_{id}$ the ignition delay \cite{heywoodbook}:
\begin{equation}
\resizebox{1\columnwidth}{!}{$
\begin{split}
\tau_{id}(\theta) &= (0.36+0.22S_p)\exp\bigg[E_a\left( \frac{1}{R_{gas}\mathcal{T}_{cyl}(\theta_{TDC})}-\frac{1}{17190}\right)\\
&\hspace{1em}\left(\frac{21.2}{\mathcal{P}_{cyl}(\theta_{TDC})-12.4}\right)^{0.63}\bigg]
\end{split}$}
\label{eq:tau_id}
\end{equation}
with $R_{gas}$ the ideal gas constant and $E_a = \frac{618840}{CN+25}$ the activation energy, function of the cetane number $CN$. $S_p$ is the mean piston speed, defined as:
\begin{equation}
S_p = \frac{2L}{60}\frac{60}{2\pi}\omega_{eng}
\end{equation}
where $L$ is the stroke. Values for temperatures and pressure at the top dead center $\theta_{TDC}$ are estimated assuming a polytropic model for compression: $\mathcal{T}_{cyl}(\theta_{TDC}) = T_Ir_c^{\gamma-1}$ and $\mathcal{P}_{cyl}(\theta_{TDC}) = P_Ir_c^{\gamma}$ ($T_I$ and $P_I$ are the intake gas temperature and pressure). Therefore, the in-cylinder temperature dynamics is obtained from the ideal gas law (deriving with respect to $\theta$):
\begin{equation}
\begin{split}
&\mathcal{T}_{cyl}(\theta)=\frac{\mathcal{P}_{cyl}(\theta)V(\theta)}{nR_{gas}},\\
&\rightarrow  \frac{d\mathcal{T}_{cyl}}{d\theta}=\frac{1}{nR_{gas}}\left[\frac{d\mathcal{P}_{cyl}}{d\theta}V(\theta)+\mathcal{P}_{cyl}(\theta)\frac{dV}{d\theta}\right] 
\end{split}
\label{eq:temperature_incyl}
\end{equation}
before injection $n=n_I$ and after injection $n=n_I + m_f/M_f$ (the number of moles is obtained integrating Equation \myref{eq:mole_rate} over one cycle).  

To compute the thermal exchange between the in-cylinder gas mixture and the walls, the Hohenberg correlation is used and the convective heat transfer coefficient is computed \cite{lakshminarayanan2010modelling}:
\begin{equation}
h_{cyl}(\theta) = 130\ \mathcal{P}_{cyl}(\theta)^{0.8}\mathcal{T}_{cyl}(\theta)^{-0.4}V(\theta)^{-0.06}(S_p+b)^{0.8}
\end{equation}
where $b$ is a tuning parameter. The convective heat exchange takes the following form:
\begin{equation}
\frac{d\mathcal{Q}_{cyl}}{d\theta}= h_{cyl}(\theta)A(\theta)(\mathcal{T}_{cyl}(\theta)-T_w)\frac{1}{\omega_{eng}}
\label{eq:heat_incyl}
\end{equation}
with $T_w$ the average cylinder wall temperature.

\begin{figure}[!t]
\centering
\includegraphics[width = \columnwidth]{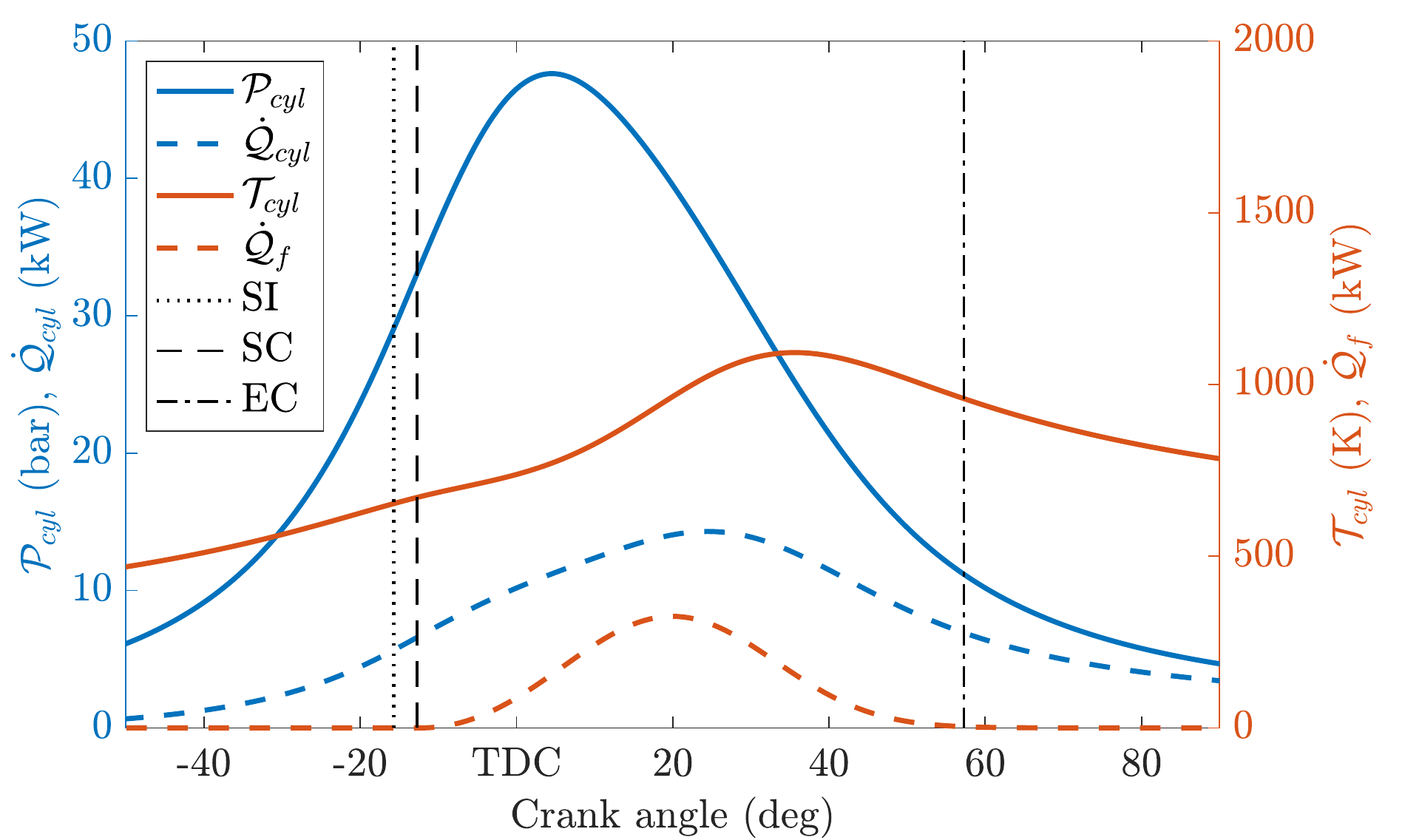}
\caption{In-cylinder pressure ($\mathcal{P}_{cyl}$), temperature ($\mathcal{T}_{cyl}$), heat transfer ($\dot{\mathcal{Q}}_{cyl}$), and heat released ($\dot{\mathcal{Q}}_{f}$) profiles are shown for compression and expansion strokes. Simulation results are obtained considering the engine working at \myunit{2000}{rpm}, with \myunit{20}{mg} of injected fuel, $\lambda$ of \myunit{3.7}, and 20\% EGR.  Definitions of acronyms EC,  SC,  SI,  and TDC are listed in the notation table, at the beginning of the paper.}
\label{fig:CA_sim}
\end{figure}

Simulation results for the engine operating at \myunit{2000}{rpm}, with 20\% EGR, are shown in Figure \ref{fig:CA_sim}. Results, in terms of $\mathcal{P}_{cyl}$, $\mathcal{T}_{cyl}$, $\dot{\mathcal{Q}}_{cyl}$, and $\dot{\mathcal{Q}}_{f}$, are in line with the literature (see \cite{heywoodbook}, Chapters 10 and 12)\footnote{To move from the crank-angle to the time domain, the relationship $d\theta = \omega_{eng}dt$ is used, thus, $\dot{\mathcal{Q}} = \frac{d\mathcal{Q}}{d\theta}\omega_{eng}$.}. Moreover, in accordance with \cite{sindhu2015real}, computing the maximum of $\mathcal{P}_{cyl}$ and $\mathcal{T}_{cyl}$ for each engine operating point leads to values confined below \myunit{90}{bar} and \myunit{1900}{K}, respectively. Similarly, the maximum of $\dot{\mathcal{Q}}_{cyl}/A$ assumes values of the order of \myunit{2}{MW/m^2} (in line with \cite{heywoodbook} and \cite{sanli2008influence}). Figures for maximum in-cylinder pressure, heat transfer, and temperature are provided in Appendix B. 

This crank-angle in-cylinder model is applicable also to spark ignition engines. In this context,  Equation \myref{eq:tau_id} should be removed and the start of combustion $\theta_{SC}$ defined with respect to the spark timing only.  It is worth mentioning that the proposed model is applicable to normal combustion events and does not capture gasoline autoignition phenomena, i.e., knock.  If abnormal combustion is a concern, readers can refer to \cite{heywoodbook,guzzellabook} for further details on the topic.

\begin{figure}[!t]
\centering
\includegraphics[width = \columnwidth]{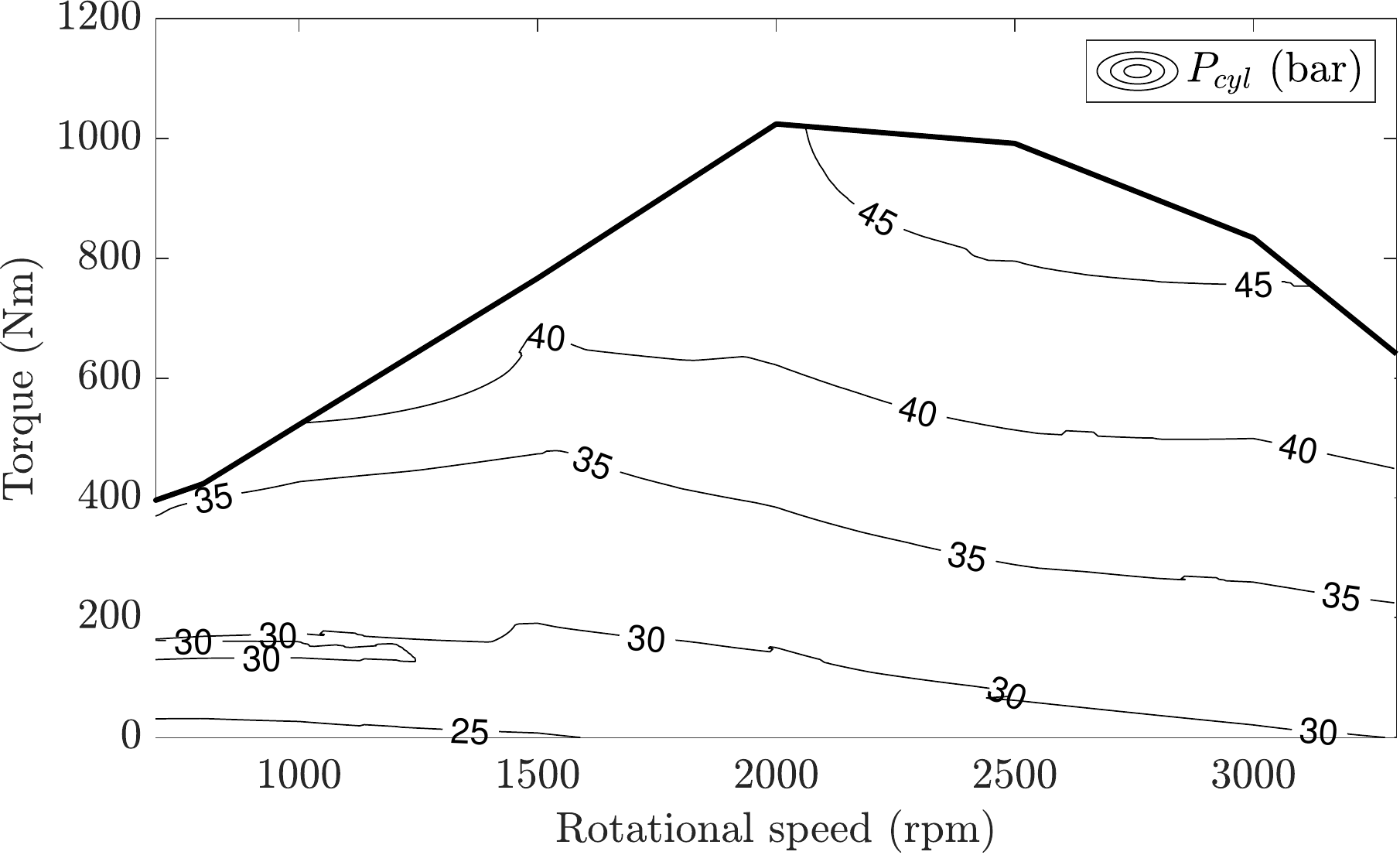}
\caption{In-cylinder average pressure for each engine operating point. The EGR rate is set to 20\%.}
\label{fig:Pcyl}
\end{figure}

\begin{figure}[!t]
\centering
\includegraphics[width = \columnwidth]{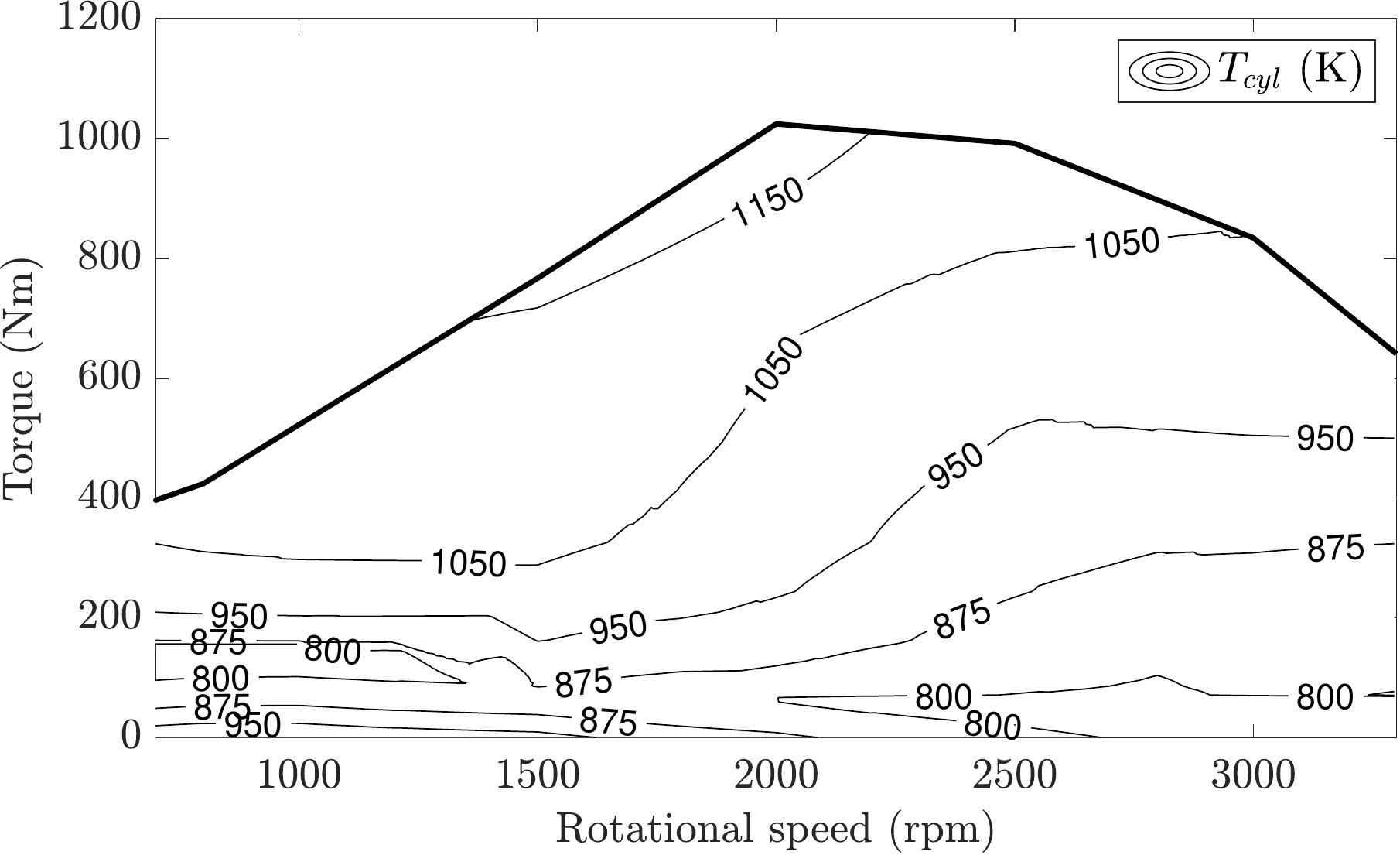}
\caption{In-cylinder average temperature for each engine operating point. The EGR rate is set to 20\%.}
\label{fig:Tcyl}
\end{figure}

\subsubsection{Mean-value model}
The model described in the previous section is employed to simulate, in the crank-angle domain, the in-cylinder pressure, heat transfer, and temperature profiles for each operating point $(\omega_{eng},\mathrm{T}_{eng})$. These simulation results are used to build static maps providing a mean-value description of the combustion process.

Given the $j$-th operating point $(\omega_{eng},\mathrm{T}_{eng})_j$, the following mean values are computed:
\begin{equation}
\bar{\mathcal{V}}_j=\frac{1}{t_2-t_1}\int_{t_1}^{t_2} \mathcal{V}_j(t)dt
\label{eq:mv_heat}
\end{equation}
where $\mathcal{V}_j = \{\mathcal{P}_{cyl,j},\dot{\mathcal{Q}}_{cyl,j},\mathcal{T}_{cyl,j}\}$ and $\bar{\mathcal{V}}_j=\{P_{cyl,j},\dot{Q}_{cyl,j},T_{cyl,j}\}$. Equation \myref{eq:mv_heat} is evaluated in the time window $(t_2-t_1)$ -- corresponding to \myunit{90}{deg} in the crank-angle domain, i.e., to the engine firing interval\footnote{The firing interval for a V8 engine is computed as $4\pi/n_{cyl}=\pi/2 \rightarrow\ $\myunit{90}{deg}.} --. Therefore, $t_1$ and $t_2$ take the following expressions:
\begin{itemize}
\item $t_1 = -(\theta_{SI}+\theta_{corr})/\omega_{eng}$;
\item $t_2 = (\theta_{EC}+\theta_{corr})/\omega_{eng}$;
\end{itemize}

\noindent with $\theta_{EC}$ the crank-angle at the end of combustion. The correction factor $\theta_{corr}=0.5(\pi/2-(\tau_{id}+\Delta\theta))$ ensures the window $t_2-t_1$ to be always corresponding to \myunit{90}{deg}. Computing the mean values over the firing interval allows to accurately capture the average heat released and transferred during the combustion process in a cylinder (in accordance with \cite{sanli2008influence}, the highest heat transfer rates take place near the TDC). This is of particular interest for the characterization of the exergy terms related to combustion irreversibilities and heat transfer in Section \ref{sec:exergy}. Recalling Assumption \ref{assum:ss_incyl}, and considering one engine cycle of duration $t_{cycle}$, $\bar{\mathcal{V}}_j$ is an invariant property of the system -- i.e., it is a constant describing the combustion process in each of the $n_{cyl}$ cylinders firing one after the other -- and can be used to describe the average behavior of the engine at the given operating point $(\omega_{eng},\mathrm{T}_{eng})_j$. 
\begin{figure}[!t]
\centering
\includegraphics[width = \columnwidth]{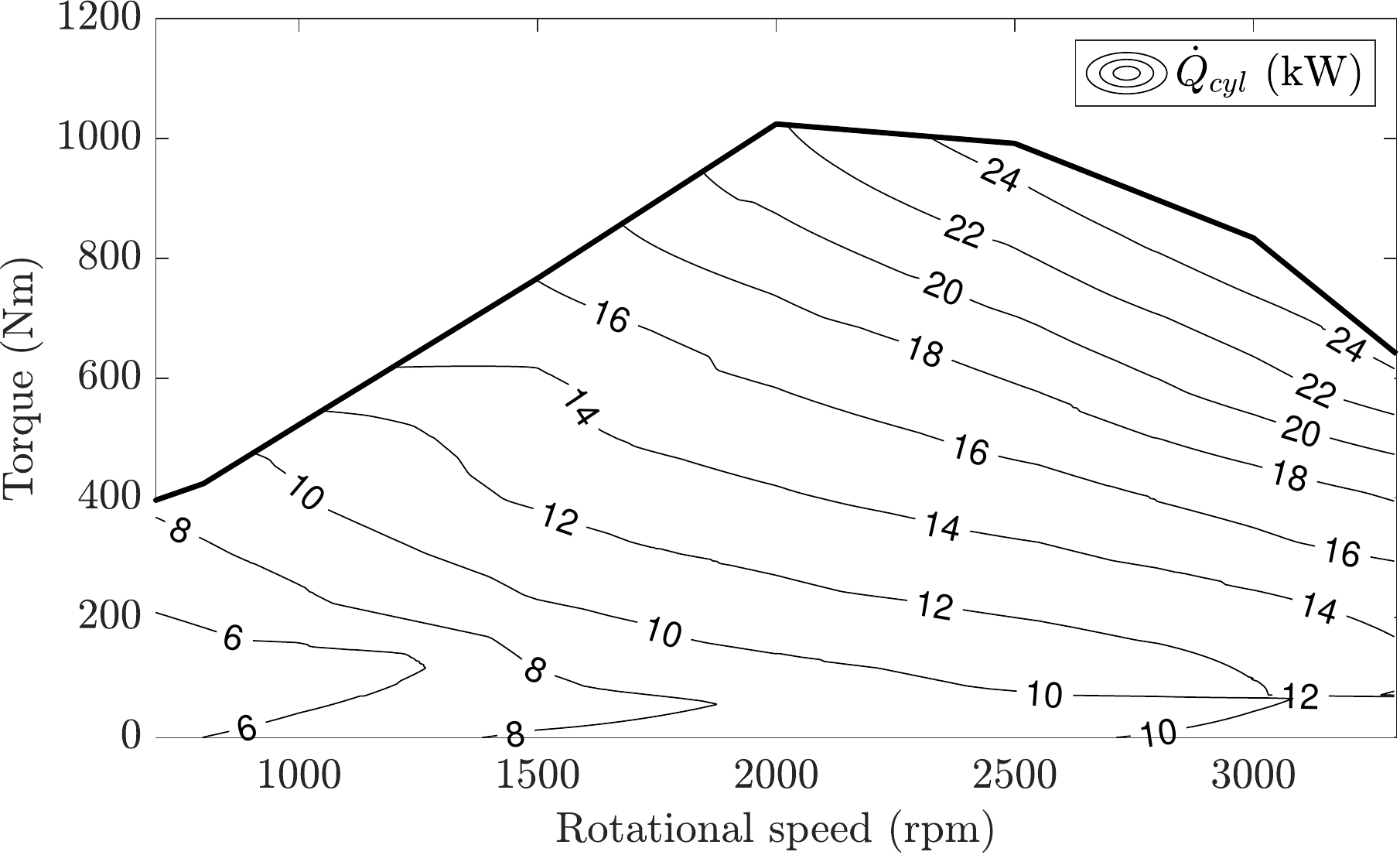}
\caption{Average heat transfer between the in-cylinder mixture and walls for each engine operating point. The EGR rate is set to 20\%.}
\label{fig:Qdcyl}
\end{figure}

Computing Equation \myref{eq:mv_heat} for each operating point $(\omega_{eng},\mathrm{T}_{eng})_j$ allows to obtain the maps in Figures \ref{fig:Pcyl}, \ref{fig:Tcyl}, and \ref{fig:Qdcyl},  useful for the formulation of the exergy balance presented in Section \ref{sec:exergy}. These maps are obtained considering an EGR rate of 20\% (see Table \ref{tab:param1}). As expected from Equations \myref{eq:pressure_incyl}, \myref{eq:temperature_incyl}, and \myref{eq:heat_incyl}, an increase in the load is related to higher fueling levels and leads to increased in-cylinder pressure, heat transfer, and temperature. Considering Figure \ref{fig:Qdcyl}, the order of magnitude of $\dot{Q}_{cyl}$ (\mytilde\myunit{10}{kW}) is in line with results shown in \cite{guzzellabook}. From now on, the following notation is used to indicate in-cylinder pressure, heat transfer, and temperature as function of the operating point:
\begin{equation}
\begin{split}
&P_{cyl}(t) = P_{cyl}(\omega_{eng}(t),\mathrm{T}_{eng}(t)),\\
&\dot{Q}_{cyl}(t) = \dot{Q}_{cyl}(\omega_{eng}(t),\mathrm{T}_{eng}(t)),\\
&T_{cyl}(t) = T_{cyl}(\omega_{eng}(t),\mathrm{T}_{eng}(t))\\
\end{split}
\end{equation}

\section{ICE exergy balance}\label{sec:exergy}\label{sec:exergy_model}
In this section, the mean-value exergy balance for the ICE is formulated.  The overall balance for this open system takes the following form:
\begin{equation}
\begin{split}
\dot{X}_{eng}(t) =&\ \underbrace{\dot{X}_{fuel,eng}(t) +  \dot{X}_{intk,eng}(t)}_{\dot{X}_{In}}+\\
&+ \underbrace{\dot{X}_{work,eng}(t) +\dot{X}_{heat,eng}(t) + \dot{X}_{exh,eng}(t)}_{\dot{X}_{Out}} +\\
&+ \underbrace{\dot{X}_{comb,eng}(t) + \dot{X}_{fric,eng}(t)}_{\dot{X}_{Dest}} + \dot{X}_{others}(t)
\end{split}
\label{eq:exergy_ice}
\end{equation}
During the engine operation, fuel and intake air are the positive exergy flows entering the control volume: $\dot{X}_{In} = \dot{X}_{fuel,eng} +  \dot{X}_{intk,eng}$. This exergy is transferred out of the control volume ($\dot{X}_{Out}$) because of mechanical work ($\dot{X}_{work,eng}$), heat transfer with the environment ($\dot{X}_{heat,eng}$), and exhaust transport ($\dot{X}_{exh,eng}$). The terms $\dot{X}_{comb,eng}$ and $\dot{X}_{fric,eng}$ indicate the portion of exergy destroyed ($\dot{X}_{Dest}$) by irreversibilities in the combustion process and piston motion. The component $\dot{X}_{others}$ accounts for all the unmodeled exergy transfer and destruction phenomena taking place in the engine, e.g., blow-by gases, losses in valves throttling, and nonuniform in-cylinder combustion (not modeled, relying on the single-zone approach). 

\begin{figure*}[!t]
\centering
\includegraphics[width = 1\columnwidth]{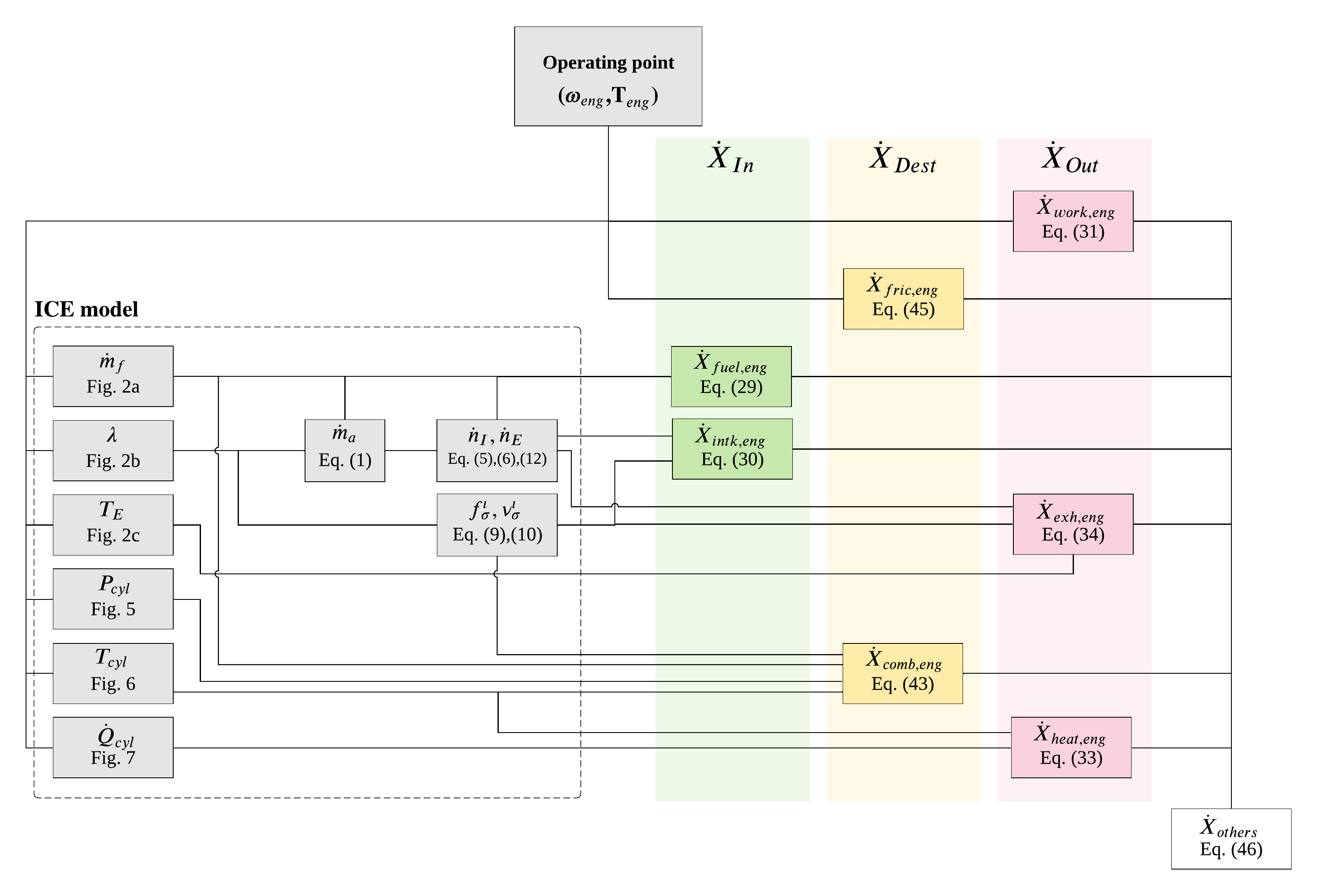}
\caption{Block diagram of the ICE exergy balance.  The green and pink colored blocks represent the exergy going into ($\dot{X}_{In}$) and out of the engine ($\dot{X}_{Out}$),  respectively.  Yellow blocks represent exergy destruction terms ($\dot{X}_{Dest}$).  Unmodeled exergy transfer and destruction phenomena are lumped in $\dot{X}_{others}$, the white block on the right. }
\label{fig:bd_exergy}
\end{figure*}

According to \cite{rakopoulos2009diesel} and Assumption \ref{assum:ss}, the engine is assumed to work always at steady-state conditions, thus, transient phenomena between two operating points are neglected. Mathematically, this condition takes the following expression:
\begin{equation}
\int_{cycle}\frac{d\mathcal{X}_{eng}}{d\theta}d\theta = 0
\label{eq:ss_xeng}
\end{equation}
with $\mathcal{X}_{eng}$ the engine exergy state in the crank-angle domain, and \myquote{$cycle$} indicating that the integral is computed along one engine cycle, i.e., \myunit{720}{deg}.  As shown in Appendix A, for the mean-value model, the following condition holds true:
\begin{equation}
\dot{X}_{eng}(t) = 0
\label{eq:ss_xeng_mv}
\end{equation}
The previous result will be employed to compute the term $\dot{X}_{others}$ in Section \ref{sec:others}. 

The overall block diagram of the ICE exergy model is shown in Figure \ref{fig:bd_exergy}.  Given an operating point $(\omega_{eng},\mathrm{T}_{eng})$ (or a sequence of operating points),  the ICE exergy terms in Equation \myref{eq:exergy_ice} are computed.  
The ICE exergy model developed in this paper has been implemented using \textsc{Matlab\&Simulink}. 

\subsection{Reference state}\label{sec:ref_envir}
Exergy terms in Equation \myref{eq:exergy_ice} are defined with respect to a reference state. In this work,  the reference state is defined to be at pressure $P_0 =\ $\myunit{1}{bar} and temperature $T_0=\ $\myunit{298.15}{K}. In accordance with \cite{razmaraexergyengine}, the atmosphere is assumed to be composed by four species $\sigma\in\mathcal{S}$ and other components (mostly argon) lumped in one term ($f_{others,0}$). The volume fraction composition of atmospheric air is as follows:
\begin{equation}
\begin{split}
f_{N_2,0} = 0&.7567,\ f_{CO_2,0} = 0.0003,\ f_{H_2O,0} = 0.0303,\\
&f_{O_2,0} = 0.2035,\ f_{others,0} = 0.0092
\end{split}
\end{equation}

\subsection{Fuel}
In accordance with \cite{razmaraexergyengine}, the availability associated to the injected fuel is computed as follows:
\begin{equation}
\begin{split}
\dot{X}_{fuel,eng}(t) &= a_f\ \dot{m}_f(t) = \\
& = \left(1.04224+0.011925\frac{x}{y}-\frac{0.042}{x}\right)LHV\ \dot{m}_f(t)
\end{split}
\label{eq:fuel_avail}
\end{equation}
where $a_f$ is the specific chemical exergy associated with the fuel. 

\subsection{Intake gas}\label{sec:intk}
The intake exergy flow is modeled considering both the chemical and physical exergies entering the system:
\begin{equation}
\begin{split}
&\dot{X}_{intk,eng}(t) = \sum_{\sigma\in\mathcal{S}} \dot{n}_{I,\sigma}(t)(\psi^I_{ch,\sigma}(t)+\psi^I_{ph,\sigma}(t)),\\
& \dot{n}_{I,\sigma}(t) = \dot{n}_{I}(t)f_\sigma^I(t),\\
& \psi^I_{ph,\sigma}(t) = \left(h_\sigma(T_I)-T_0s_\sigma(T_I)\right) - \left(h_\sigma^\star(T_0)-T_0s_\sigma^\star(T_0)\right),\\
& \psi^I_{ch,\sigma}(t) = R_{gas}T_0\lnp{\frac{f_\sigma^\star(t)}{f_{\sigma,0}}}=  R_{gas}T_0\lnp{\frac{f_\sigma^I(t)}{f_{\sigma,0}}}
\end{split}
\label{eq:intk_avail}
\end{equation}
with $\psi^I_{ph,\sigma}$ the physical exergy, modeling the work potential between the current system state (at $T_I$ and $P_I$) and the restricted state, and $\psi^I_{ch,\sigma}$ the chemical exergy, accounting for the different chemical composition between the restricted (i.e., the intake manifold composition) and reference state. The term $\dot{n}_{I,\sigma}$ indicates the fraction of the total intake molar flow rate for a particular species $\sigma$. 

\subsection{Mechanical work}
A portion of the availability introduced in the engine, modeled as in Equations \myref{eq:fuel_avail} and \myref{eq:intk_avail}, is converted into useful mechanical work (brake work):
\begin{equation}
\dot{X}_{work,eng}(t) = -\mathrm{T}_{eng}(t)\omega_{eng}(t) = - P_{eng}(t)
\end{equation}
The percentage of the fuel availability converted into mechanical work is the engine efficiency according to the second law of thermodynamics, i.e.,
\begin{equation}
\varepsilon(t) = \frac{{X}_{work,eng}(t)}{{X}_{fuel,eng}(t)}
\label{eq:ex_eff_2nd}
\end{equation}
where ${X}_{work,eng}$ and ${X}_{fuel,eng}$ are obtained integrating over time  the corresponding exergy rate terms.
 
\subsection{Heat exchange}
The exergy transfer, due to heat exchange between the in-cylinder mixture and the cylinder's walls, is computed relying on Figures \ref{fig:Tcyl} and \ref{fig:Qdcyl}:
\begin{equation}
\dot{X}_{heat,eng}(t) = \left(1-\frac{T_0}{T_{cyl}(t)}\right)\left(-\dot{Q}_{cyl}(t)\right)
\label{eq:heat_tran_ex}
\end{equation}
where the term inside brackets is the Carnot heat engine efficiency. The temperature $T_{cyl}$, at which the thermal exchange occurs, must be higher than the reference state temperature $T_0$. Also, as the temperature $T_{cyl}$ decreases, the exergy transfer is reduced because the Carnot efficiency, function of the ratio between the low- and high- temperature reservoirs, is reduced. The minus before $\dot{Q}_{cyl}$ models the direction of the heat transfer: from inside the engine to the environment.

\subsection{Exhaust gas}
The exhaust gas exergy transfer is modeled in accordance with Section \ref{sec:intk}, i.e., considering the chemical and physical exergy fluxes for each species $\sigma$:
\begin{equation}
\begin{split}
&\dot{X}_{exh,eng}(t) = -\sum_{\sigma\in\mathcal{S}} \dot{n}_{E,\sigma}(t)(\psi^E_{ch,\sigma}(t)+\psi^E_{ph,\sigma}(t)),\\
& \dot{n}_{E,\sigma}(t) = \dot{n}_{E}(t)f_\sigma^E(t),\\
& \psi^E_{ph,\sigma}(t) = \left(h_\sigma(T_E(t))-T_0s_\sigma(T_E(t))\right) - \left(h_\sigma^\star(T_0)-T_0s_\sigma^\star(T_0)\right),\\
& \psi^E_{ch,\sigma}(t) = R_{gas}T_0\lnp{\frac{f_\sigma^\star(t)}{f_{\sigma,0}}}=  R_{gas}T_0\lnp{\frac{f_\sigma^E(t)}{f_{\sigma,0}}}
\end{split}
\label{eq:exh_avail}
\end{equation}
with $\psi^E_{ph,\sigma}$ the physical exergy, modeling the work potential between the current system state (at $T_E$ and $P_E$) and the restricted state, and $\psi^E_{ch,\sigma}$ the chemical exergy, accounting for the different chemical composition between the restricted (i.e., the exhaust manifold composition) and reference state. The term $\dot{n}_{E,\sigma}$ indicates the fraction of the total exhaust molar flow rate for a particular species $\sigma$. 

\subsection{Combustion irreversibilities}
Combustion irreversibilities are function of the reaction rate and of the variations of the chemical potential between reactants and products. Thus, the entropy generation in chemical reactions ($\dot{S}_{gen,eng}$), formulated as in \cite{prigogine1967introduction}, is employed to model the exergy destruction due to combustion irreversibilities:
\begin{equation}
\resizebox{1\columnwidth}{!}{$
\begin{split}
\dot{X}&_{comb,eng}(t)=-T_{0}\dot{S}_{gen,eng}(t)=\\
&=\frac{T_0}{T_{cyl}(t)}\bigg\lbrace \underbrace{-\nu_{f}\mu_f + \sum_{\sigma\in\mathcal{S}} \big[-\nu^I_\sigma(t)\mu^I_\sigma(t)}_{Reactants} + \underbrace{\nu^E_\sigma(t)\mu^E_\sigma(t)}_{Products}\big]\bigg\rbrace\dot{\xi}(t)
\end{split}$}
\label{eq:irrex}
\end{equation}
where $I$ and $E$ indicate the gaseous species entering -- reactants --  and leaving -- products -- the engine and $\nu$ are the stoichiometric coefficients of the combustion reaction \myref{eq:comb_react}. The chemical reaction is assumed to take place at the average in-cylinder temperature $T_{cyl}$ (Figure \ref{fig:Tcyl}). Thus, for each gaseous species $\sigma$ in $\iota\in\{I,E\}$, the chemical potential $\mu$ is defined as:
\begin{equation}
\mu_\sigma^{\iota}(t) = g_\sigma(t) + R_{gas}T_{cyl}(t)\lnp{\frac{f_\sigma^{\iota}(t)P_{cyl}(t)}{P_0}}
\label{eq:chempot}
\end{equation}
with $g_\sigma(t)=h_\sigma(T_{cyl}(t))-T_{cyl}(t)s_\sigma(T_{cyl}(t))$ the Gibbs free energy. It is worth to mention that, the entropy generation $\dot{S}_{gen,eng}$, is function of $T_{cyl}$: the higher the temperature, the higher the combustion efficiency and the lower the entropy generation.
\noindent In Equation \myref{eq:irrex}, reactants are consumed and converted into products. The extent of reaction $\xi$ models the reactants to products conversion rate:
\begin{equation}
\dot{\xi}(t) = \frac{1}{\nu_{\scriptscriptstyle{f}}}\dot{n}_{\scriptscriptstyle{C_xH_y}}(t) = \frac{1}{\nu^I_\sigma(t)}\dot{n}_{I,\sigma}(t) = \frac{1}{\nu^E_\sigma(t)}\dot{n}_{E,\sigma}(t)
\end{equation}
with $\dot{n}_{\scriptscriptstyle{C_xH_y}}$ the reacted fuel. Assuming the fuel injected $\dot{n}_f$, for a given operating point, to completely react with the intake gas (see Assumption \ref{assum:fuelburn}),  and recalling that $\nu_f =1$, the following equality holds:
\begin{equation}
\dot{\xi}(t) = \dot{n}_{C_xH_y}(t)=\dot{n}_f(t)
\label{eq:extent_1}
\end{equation}
which leads to:
\begin{equation}
\dot{n}_{I,\sigma}(t) =\nu^I_{\sigma}\dot{n}_f(t), \hspace{0.5em} \dot{n}_{E,\sigma}(t) =\nu^E_{\sigma}\dot{n}_f(t)
\label{eq:extent_2}
\end{equation}

\noindent To compute the hydrocarbons chemical potential $\mu_f=g_f$ (the fuel is considered in its liquid state), the expression of the specific fuel exergy $a_f$ is recalled. In accordance with \cite{bejan1999thermodynamic,dincer2012exergy}, considering a simplified combustion reaction scheme:
\begin{equation}
C_xH_x + \left(x+\frac{y}{4}\right)O_2\ \contour{black}{$\rightarrow$}\ xCO_2+\frac{y}{2}H_2O
\end{equation}
and assuming the fuel to enter the system at $P_0$ and $T_0$, the work that can be obtained through the reaction of $C_xH_y$ is shown in Equation \myref{eq:gf_1}. 
\begin{figure*}[!ht]
\begin{equation}
\resizebox{1\textwidth}{!}{$
a_f=\left(1.04224 + 0.011925\frac{x}{y} - \frac{0.042}{x}\right)LHV=\bigg[g_f+\overbrace{\left(x+\frac{y}{4}\right)g_{O_2} - xg_{CO_2}-\frac{y}{2}g_{H_2O}\bigg]_{(T_0,P_0)}+x\ \mathrm{ex}_{CO_2}^{ch} + \frac{y}{2}\mathrm{ex}_{H_2O}^{ch} -\left(x+\frac{y}{4}\right)\mathrm{ex}_{O_2}^{ch}}^{\boldsymbol\alpha}$}
\label{eq:gf_1}
\end{equation}
\end{figure*}
\noindent Therefore, inverting Equation \myref{eq:gf_1}, the fuel Gibbs free energy is obtained:
\begin{equation}
g_f = \left(1.04224 + 0.011925\frac{x}{y} - \frac{0.042}{x}\right)LHV - \boldsymbol\alpha
\label{eq:gf_2}
\end{equation}
According to \cite{rakopoulosexergyengine}, modifications in the  fuel temperature lead to negligible variations in the fuel availability $a_f$, thus, Equation \myref{eq:gf_2} can be considered a good estimate of the fuel Gibbs free energy at $T_{cyl}$ and $P_{cyl}$.

Starting from Equation \myref{eq:irrex}, while relying on Equations \myref{eq:chempot}, \myref{eq:extent_1}, \myref{eq:extent_2}, and \myref{eq:gf_2}, the final expression for $\dot{X}_{comb,eng}$ is given in Equation \myref{eq:allcombex}. 

\begin{figure*}[!ht]
\begin{equation}
\resizebox{1\textwidth}{!}{$
\begin{split}
\dot{X}_{comb,eng}(t)&=\frac{T_0}{T_{cyl}(t)}\bigg\lbrace -\nu_{f}\mu_f + \sum_{\sigma\in\mathcal{S}} \big[-\nu^I_\sigma(t)\mu^I_\sigma(t) + \nu^E_\sigma(t)\mu^E_\sigma(t)\big]\bigg\rbrace\dot{n}_f(t) =\\ 
& = -\frac{T_0}{T_{cyl}(t)}\Bigg[g_f + g_{N_2}(t)\left(\nu_{N_2}^I(t) -\nu_{N_2}^E(t)\right) + g_{CO_2}(t)\left(\nu_{CO_2}^I -\nu_{CO_2}^E\right) + g_{H_2O}(t)\left(\nu_{H_2O}^I -\nu_{H_2O}^E\right) + g_{O_2}(t)\left(\nu_{O_2}^I(t) -\nu_{O_2}^E(t)\right) + \\
& \hspace{5.5em}+\nu_{N_2}^I(t)R_{gas}T_{cyl}(t)\lnp{\frac{f_{N_2}^I(t)P_{cyl}(t)}{P_0}} + \nu_{CO_2}^IR_{gas}T_{cyl}(t)\lnp{\frac{f_{CO_2}^I(t)P_{cyl}(t)}{P_0}} +\\
& \hspace{5.5em}+ \nu_{H_2O}^IR_{gas}T_{cyl}(t)\lnp{\frac{f_{H_2O}^I(t)P_{cyl}(t)}{P_0}}  + \nu_{O_2}^I(t)R_{gas}T_{cyl}(t)\lnp{\frac{f_{O_2}^I(t)P_{cyl}(t)}{P_0}} -\\
& \hspace{5.5em} -\nu_{N_2}^E(t)R_{gas}T_{cyl}(t)\lnp{\frac{f_{N_2}^E(t)P_{cyl}(t)}{P_0}} - \nu_{CO_2}^ER_{gas}T_{cyl}(t)\lnp{\frac{f_{CO_2}^E(t)P_{cyl}(t)}{P_0}} -\\
& \hspace{5.5em}- \nu_{H_2O}^ER_{gas}T_{cyl}(t)\lnp{\frac{f_{H_2O}^E(t)P_{cyl}(t)}{P_0}}  - \nu_{O_2}^E(t)R_{gas}T_{cyl}(t)\lnp{\frac{f_{O_2}^E(t)P_{cyl}(t)}{P_0}}\Bigg]\dot{n}_f(t) = \\
& = -\frac{T_0}{T_{cyl}(t)}\Bigg\{ g_f -xg_{CO_2}(t) -\frac{y}{2}g_{H_2O}(t) + \left(x+\frac{y}{4} \right)g_{O_2}(t) + \frac{\lambda(t)}{1-x_{EGR}}\left(x+\frac{y}{4}\right)3.76R_{gas}T_{cyl}(t)\lnp{\frac{f_{N_2}^I(t)}{f_{N_2}^E(t)}} + \\
& \hspace{5.5em} + R_{gas}T_{cyl}(t)\sum_{\sigma\in\mathcal{S}\setminus\{N_2\}}\Bigg[ \nu_{\sigma}^I(t)\lnp{\frac{f_{\sigma}^I(t)P_{cyl}(t)}{P_0}} - \nu_{\sigma}^E(t)\lnp{\frac{f_{\sigma}^E(t)P_{cyl}(t)}{P_0}} \Bigg] \Bigg\}\dot{n}_f(t)
\end{split}$}\label{eq:allcombex}
\end{equation}
\end{figure*}

\subsection{Frictions}\label{sec:ex_fric}
According to \cite{heywoodbook}, the engine frictions, in terms of components motion and turbulent dissipation, are modeled according to the friction mean effective pressure (FMEP):
\begin{equation}
\text{FMEP}(t) = 1000\left[C_1 + C_2\omega_{eng}(t) + C_3S_p(t)^2\right]
\end{equation}
where $C_1$, $C_2$, and $C_3$ are identified coefficients. At last, the corresponding exergy destruction term is computed as follows:
\begin{equation}
\begin{split}
\dot{X}_{fric,eng}(t) &= - \frac{1}{t_{cycle}(t)}\text{FMEP}(t)\ V_{d,tot} =\\
&= - \frac{\omega_{eng}(t)}{4\pi}\text{FMEP}(t)\ V_{d,tot}
\end{split}
\label{eq:ex_frictions}
\end{equation}
with $t_{cycle}(t)=4\pi/\omega_{eng}(t)$ the engine cycle time for a given operating point and considering a four stroke engine.

\subsection{Others}\label{sec:others}
Recalling Equation \myref{eq:ss_xeng_mv}, the exergy term for the in-cylinder exergy transfer and destruction unmodeled phenomena is computed as follows:
\begin{equation}
\begin{split}
\dot{X}_{others}(t) =-&\big[\dot{X}_{fuel,eng}(t) +  \dot{X}_{intk,eng}(t) +\\
&+  \dot{X}_{work,eng}(t) + \dot{X}_{heat,eng}(t) + \dot{X}_{exh,eng}(t) + \\
&+ \dot{X}_{comb,eng}(t)+\dot{X}_{fric,eng}(t)\big]
\end{split}
\label{eq:exergy_others}
\end{equation}

\section{Results}\label{sec:results} 
The exergy model developed and presented step-by-step in Section \ref{sec:exergy_model} is used to assess the ICE exergetic behavior.  Firstly, each engine operating point is analyzed at steady-state according to the block diagram in Figure \ref{fig:bd_exergy} and static maps describing the exergy transfer and destruction phenomena are obtained. Secondly, the exergy analysis is shown for a sequence of engine operating points and the effect of different EGR rates is described. 

\subsection{Steady-state analysis}
In this section,  steady-state operating points characterizations are obtained through simulation using the parameters listed in Table \ref{tab:param1}.
\begin{figure}[!t]
\centering
\includegraphics[width = \columnwidth]{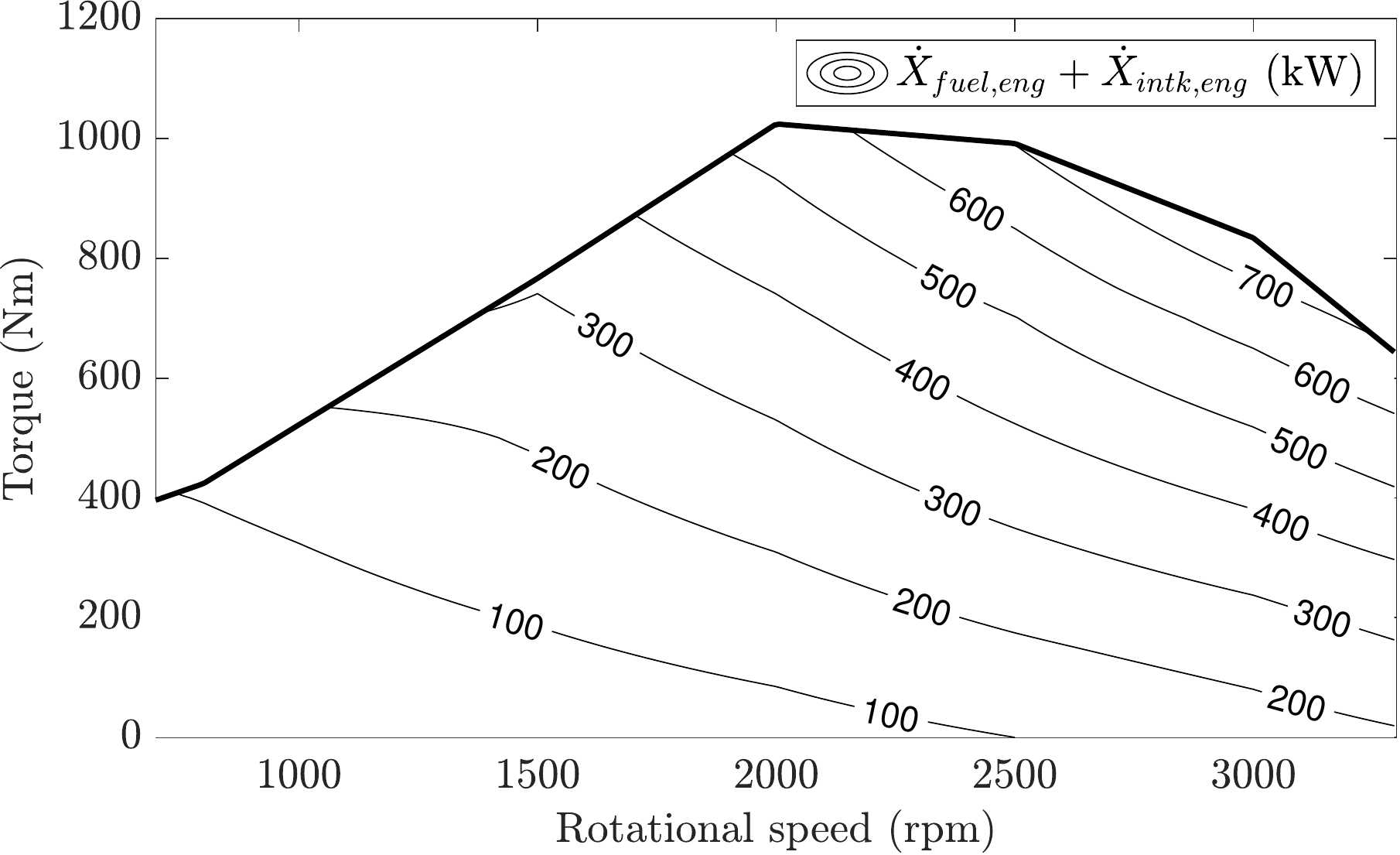}
\caption{Availability introduced in the engine by the injected fuel and the inlet manifold air: $\dot{X}_{fuel,eng}+\dot{X}_{intk,eng}$.}
\label{fig:aval_input}
\end{figure}

\begin{figure*}[!t]
\centering
\subfloat[$\dot{X}_{work,eng}$]{\includegraphics[width=0.49\textwidth]{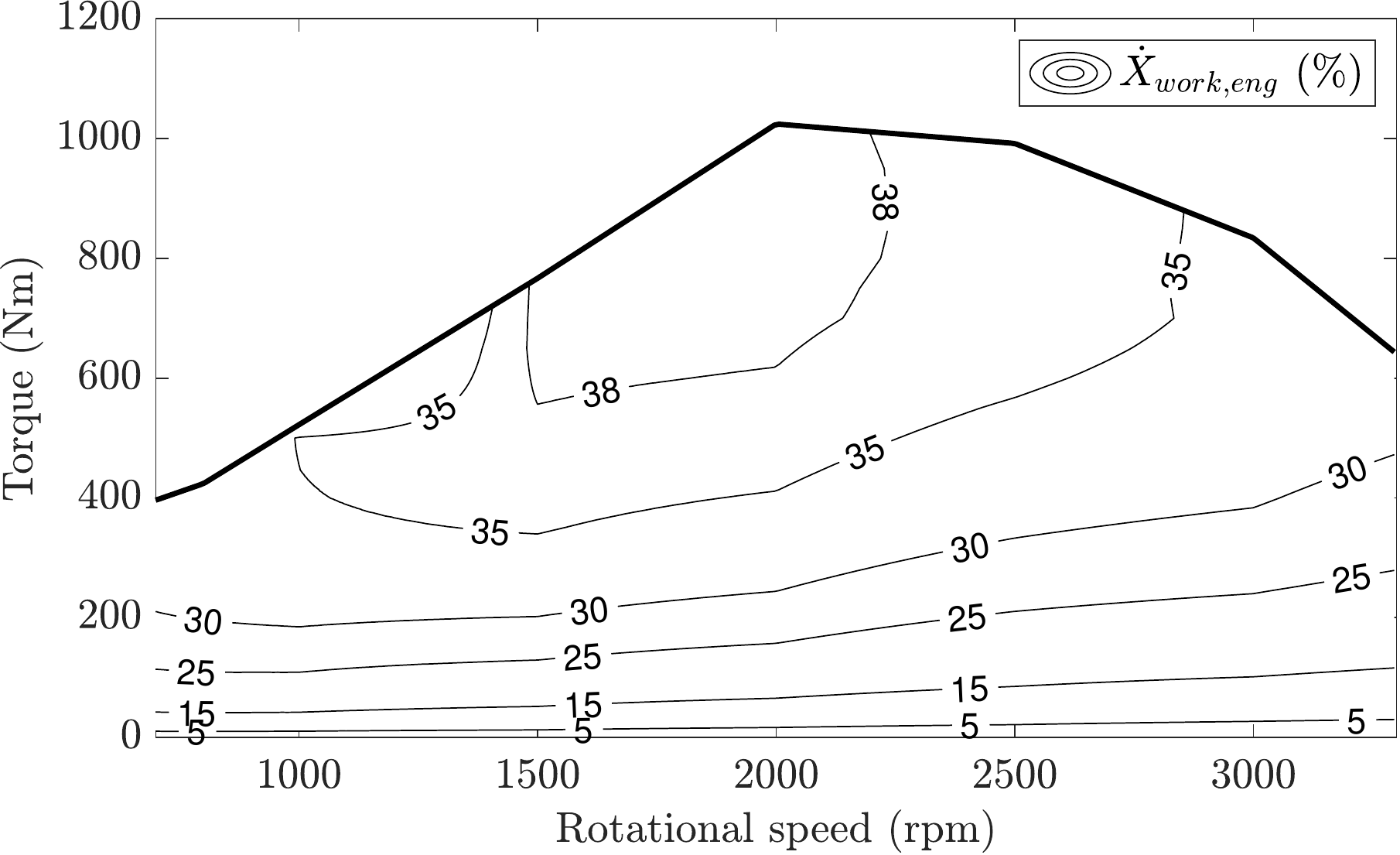}}\hspace{0.2em}
\subfloat[$\dot{X}_{heat,eng}$]{\includegraphics[width=0.49\textwidth]{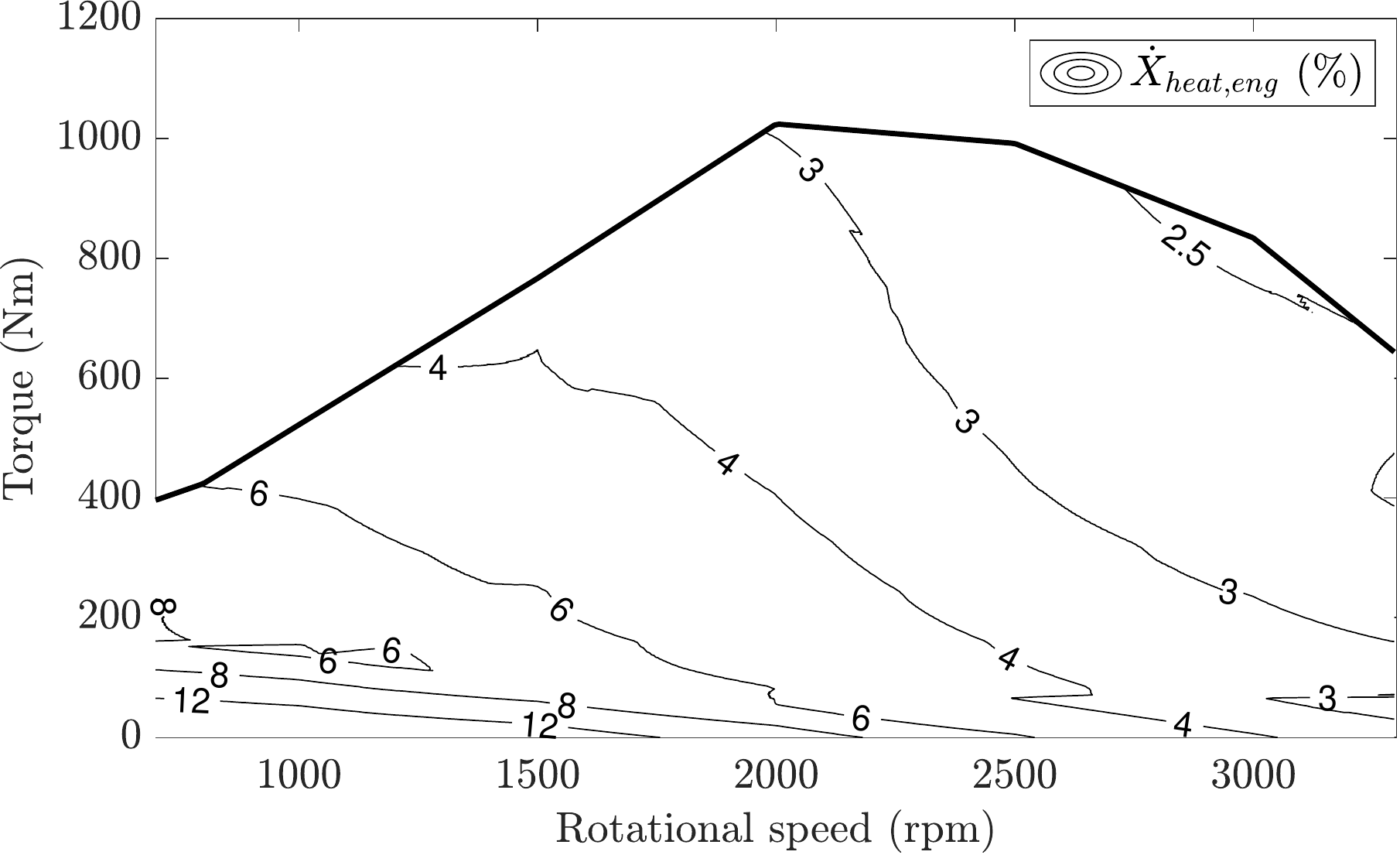}}\\
\subfloat[$\dot{X}_{exh,eng}$]{\includegraphics[width=0.49\textwidth]{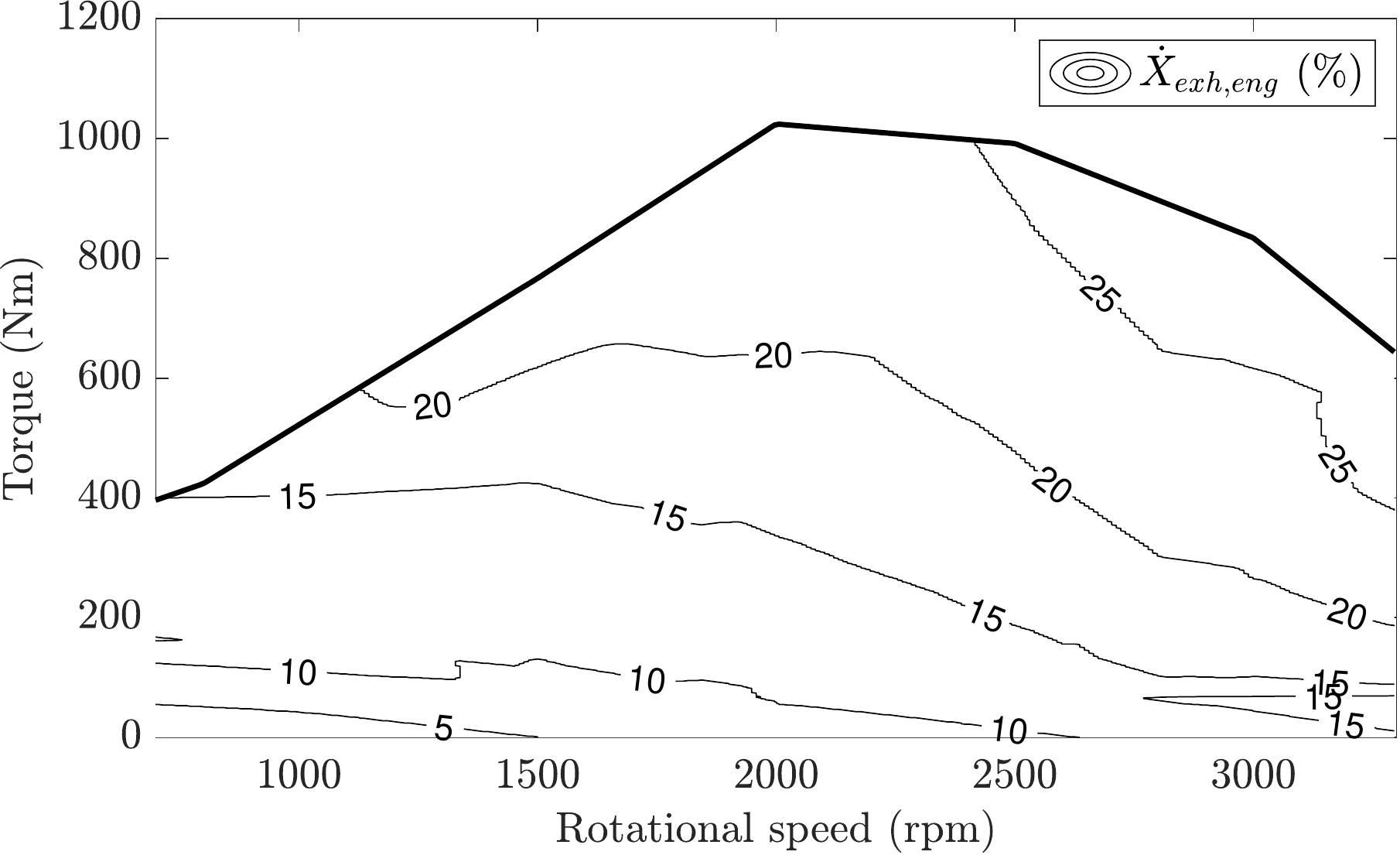}}\hspace{0.2em}
\subfloat[$\dot{X}_{comb,eng}$]{\includegraphics[width=0.49\textwidth]{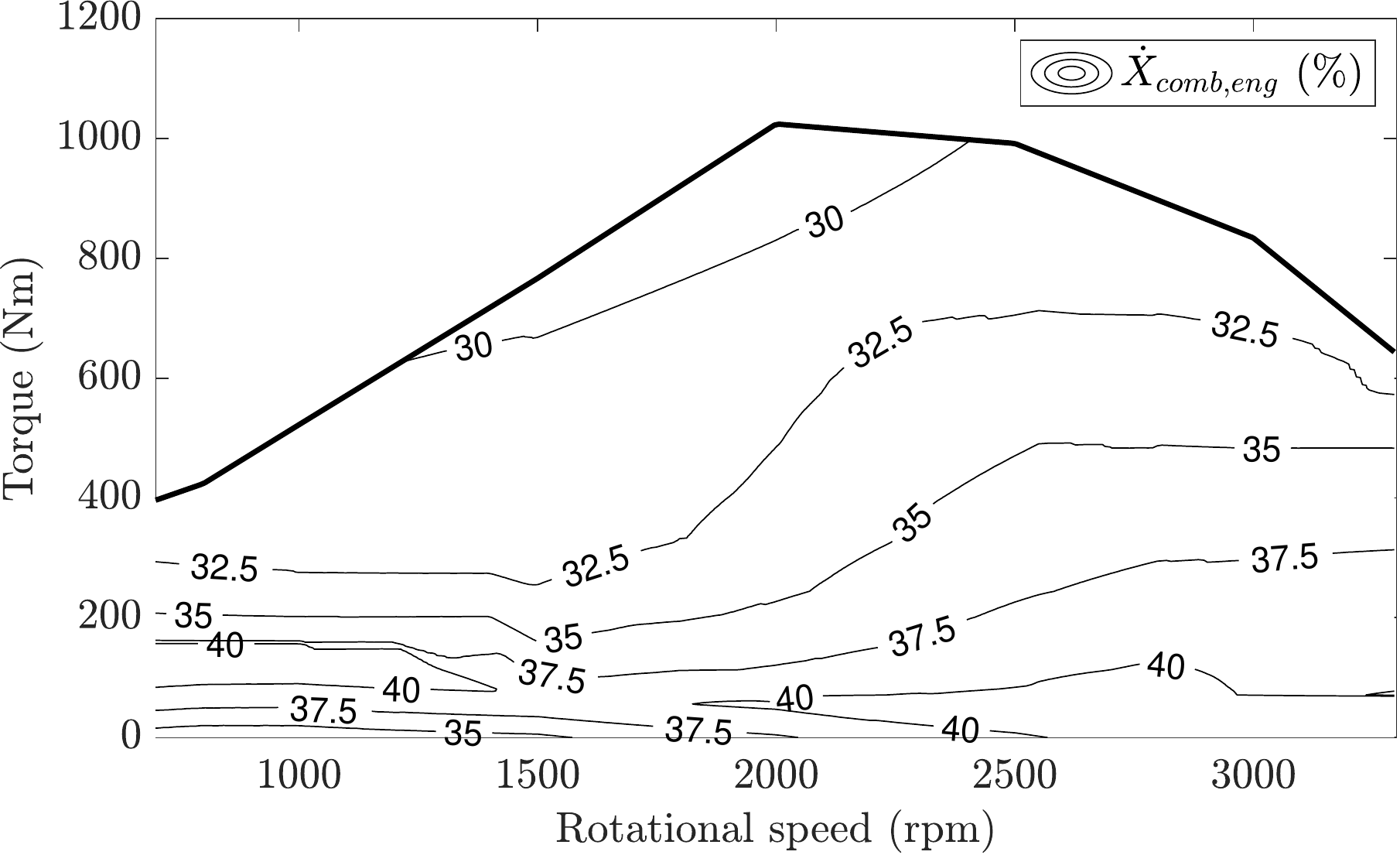}}\\
\subfloat[$\dot{X}_{fric,eng}$]{\includegraphics[width=0.49\textwidth]{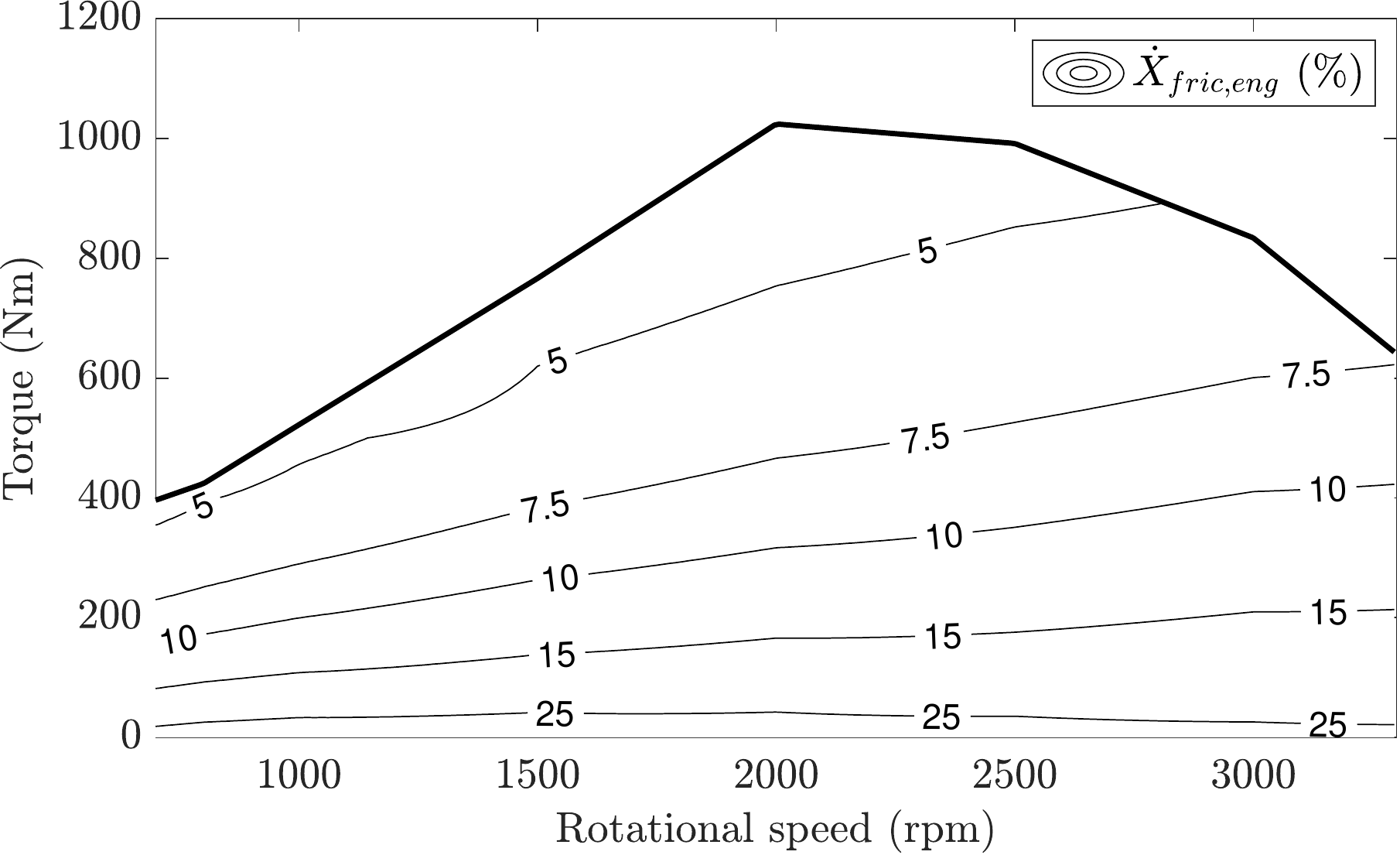}}\hspace{0.2em}
\subfloat[$\dot{X}_{others}$]{\includegraphics[width=0.49\textwidth]{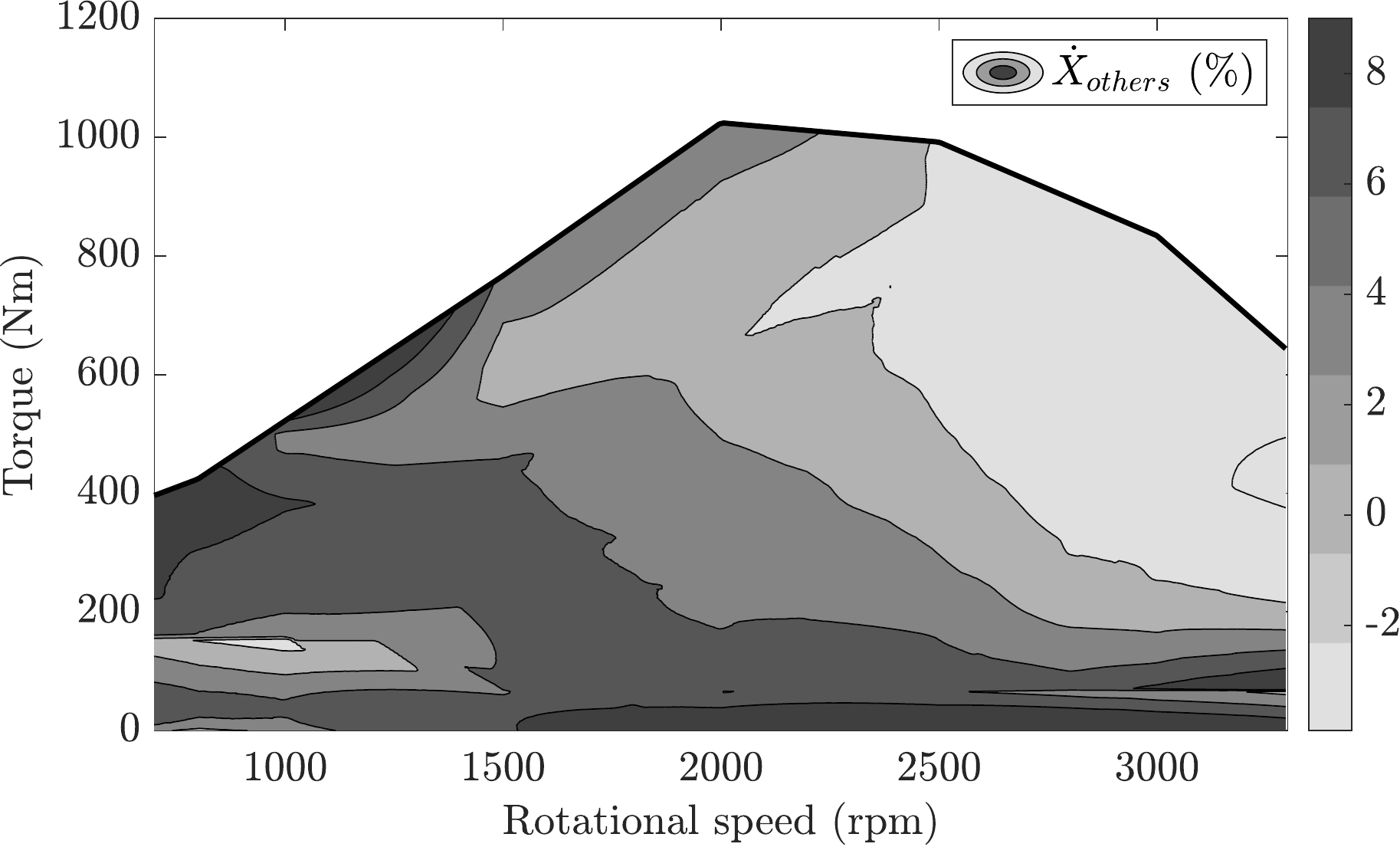}}\\
\caption{Percentage contribution of the different exergy transfer and destruction terms with respect to $\dot{X}_{fuel,eng}+\dot{X}_{intk,eng}$. Considering $\dot{X}_{others}$,  shades of gray are used to improve readability.}
\label{fig:aval_output}
\end{figure*}

The availability $\dot{X}_{In}$ introduced in the engine, and given by the summation of the injected fuel and the intake air exergies, is shown in the map of Figure \ref{fig:aval_input}. The intake air enters the cylinder at a temperature $T_I$, close to the reference state temperature $T_0$.  In accordance with \cite{sayin2016energy}, the associated exergy contribution is almost negligible and accounts for at most 1\% of the total input exergy. The remaining 99\% comes from the fuel availability defined in Equation \myref{eq:fuel_avail}. According to Section \ref{sec:exergy},  a portion of $\dot{X}_{In}$ is transferred outside the system by mechanical work, heat exchange with the cylinder's walls, and exhaust gas flow ($\dot{X}_{Out}$). The remaining input availability is destroyed in two principal ways ($\dot{X}_{Dest}$): frictions and combustion irreversibilities. To assess the relative contribution of each exergy transfer and destruction term, the following relationship is introduced:
\begin{equation}
\dot{X}_{k}\ (\%) = \frac{\dot{X}_{k}}{\dot{X}_{fuel,eng}+\dot{X}_{intk,eng}}\times 100
\label{eq:rel_contr_ex}
\end{equation}
with $\dot{X}_k\in\mathcal{K}=\{\dot{X}_{fuel,eng},\dot{X}_{intk,eng},\dot{X}_{work,eng},$ $\dot{X}_{heat,eng},\dot{X}_{exh,eng},\dot{X}_{comb,eng},\dot{X}_{fric,eng},\dot{X}_{others}\}$. 

Increasing the load leads to higher in-cylinder pressure, heat transfer, and temperature (Figures \ref{fig:Pcyl}, \ref{fig:Tcyl}, and \ref{fig:Qdcyl}). In this scenario, the percentage contribution related to combustion reactions decreases: from 40\% at low loads to around 30\% (Figure \ref{fig:aval_output}d). This is reasonable since, at higher in-cylinder temperatures, the combustion becomes more efficient and entropy generation is reduced. Conversely, the efficiency $\varepsilon$  -- defined in Equation \myref{eq:ex_eff_2nd} -- increases, leading to higher availability transfer due to mechanical work generation (up to 38\%), as shown in Figure \ref{fig:aval_output}a. An increased load means also increased exhaust gas temperatures ($T_E$), leading to higher availability transfer due to exhaust gases (Figure \ref{fig:aval_output}c).

Recalling Equation \myref{eq:heat_tran_ex}, an increasing load increases the exergy transfer due to heat exchange (the Carnot efficiency increases). However, compared to the other exergy transfer and destruction terms, the relative contribution becomes smaller. This effect is emphasized at high engine rotational speeds because of the lower available time for heat exchanges. As shown in Figure \ref{fig:aval_output}b, the contribution related to heat transfer goes from 12\%, at low loads and speed, to 2.5\%, at high loads and speed.

According to Section \ref{sec:ex_fric}, the exergy term related to frictions is an increasing function of the engine rotational speed only. At a given speed, increasing the load leads to no variation of the friction exergy term, thus, its relative contribution decreases, reaching 5\% (Figure \ref{fig:aval_output}e). At low loads, a contribution of 25\% is obtained: this is in line with the fact that the efficiency $\varepsilon$ is reduced.
\renewcommand{\arraystretch}{1.3}
\begin{table}[t!]
	\centering
	\begin{center}
		\caption{Comparison of exergy transfer and destruction phenomena at \myunit{1400}{rpm} and \myunit{258}{Nm}. In \cite{canakci2006energy}, terms $\dot{X}_{comb,eng}$, $\dot{X}_{fric,eng}$, and $\dot{X}_{others}$ are lumped together.}
		\label{tab:comparison_literauore}
		\resizebox{0.9\columnwidth}{!}{
		\begin{tabular}{cccc}
			\textbf{Exergy term} (\%) & \textit{Ref. \cite{canakci2006energy}} & \begin{tabular}{c} \textit{Engine maps}\vspace{-0.3em}\\ \textit{(Figure \ref{fig:aval_output})} \end{tabular} \\ \hline
			$\dot{X}_{work,eng}$ &  34.4 &  32.5\\
			$\dot{X}_{heat,eng}$ &  5.7 & 6.0 \\
			$\dot{X}_{exh,eng}$ &  13.6 & 12.2 \\
			\begin{tabular}{c} $\dot{X}_{comb,eng}+\dot{X}_{fric,eng}+$\vspace{-0.3em}\\ $+\dot{X}_{others}$\end{tabular} & 46.3  & 49.3 \\
			\hline 
		\end{tabular}}
	\end{center}
\end{table}

The term $\dot{X}_{others}$ accounts for all the unmodeled exergy transfer and destruction phenomena. In accordance with results shown in \cite{razmaraexergyengine,rakopoulosexergyengine}, this term accounts for 8 to 2\% (Figure \ref{fig:aval_output}f). On average, a contribution of 3.5\% is obtained.

In Table \ref{tab:comparison_literauore}, simulation data for the engine operating at \myunit{1400}{rpm} and \myunit{258}{Nm} are compared to results from \cite{canakci2006energy} obtained for a turbocharged diesel engine with a cylinder displacement of 1 liter (similar to the ICE considered in this work, for which the cylinder displacement is 0.8 liters). Given the different engines, and the dissimilar underlying control strategies, the comparison is not strictly proper. However, it provides useful insights showing that the outcomes of the proposed modeling strategy are in line with the results in \cite{canakci2006energy}, where only one operating point is analyzed.  Values in Figure \ref{fig:aval_output} are also comparable to the ranges reported in \cite{rakopoulosexergyengine} for diesel engines, i.e., the indicated work rate ($\dot{X}_{work,eng}+\dot{X}_{fric,eng}$) accounts for 40-45\%, the heat transfer term for \mytilde10\%, the exhaust gas for 10-20\%, and the combustion irreversibilities for \mytilde25\% (at full load).  On the contrary to approaches proposed in the literature,  the model developed in this paper and the static maps of Figure \ref{fig:aval_output} allow to characterize all the engine operating points and not, as has been done so far, a restricted set of operating conditions.


\subsection{Analysis over a sequence of operating points}
\begin{figure}[!tb]
\centering 
\includegraphics[width = 0.99\columnwidth]{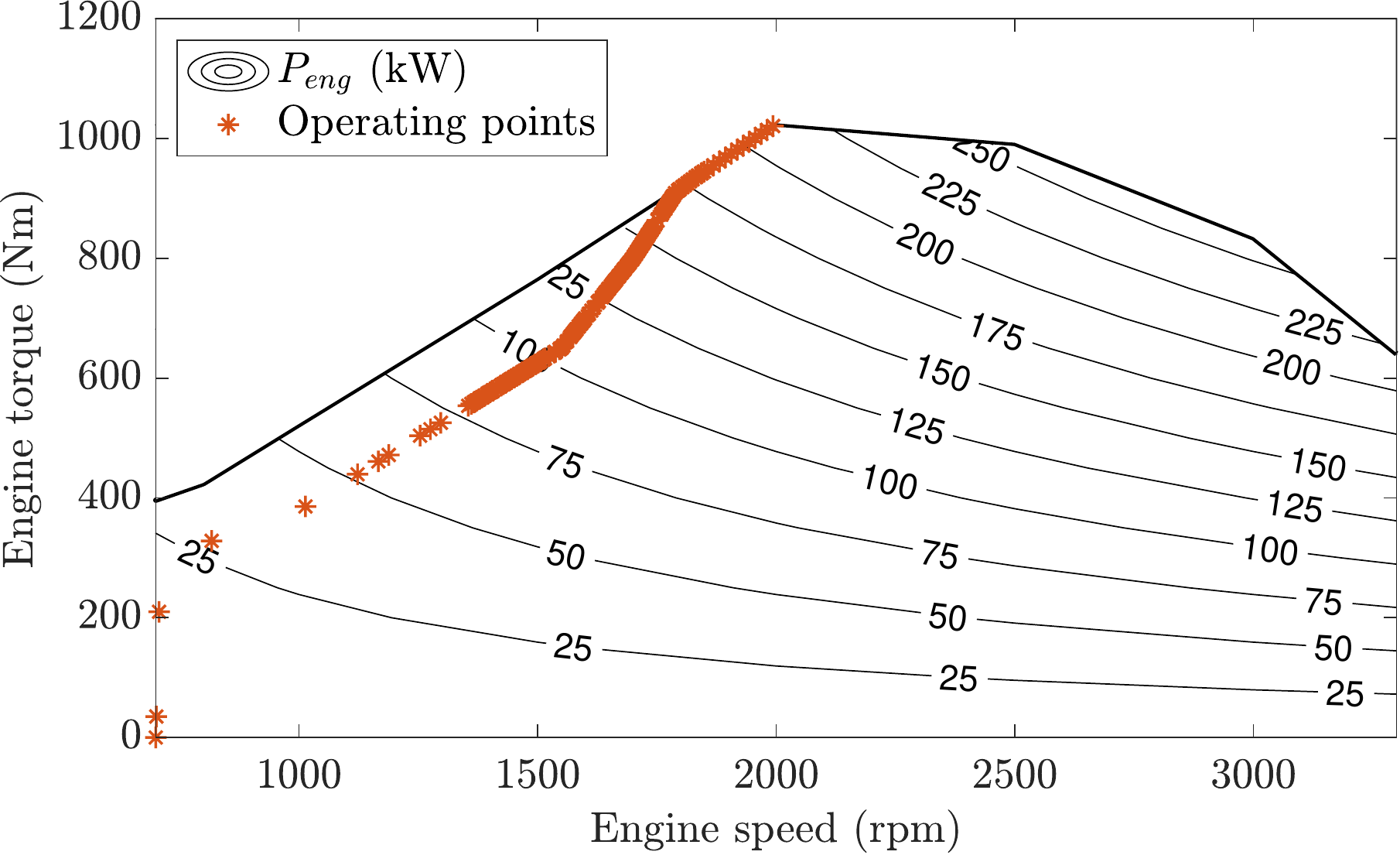}
\caption{Sequence of engine operating points and corresponding engine powers.  In this figure, $P_{eng}\in[0,214]\myspace{0.1}\mathrm{kW}$.}
\label{fig:pmp_hev}	
\end{figure}

The exergy analysis is performed considering the sequence of engine operating points depicted in Figure \ref{fig:pmp_hev},  spanning engine powers ($P_{eng}$) in the range $[0,214]\myspace{0.1}\mathrm{kW}$.  Each operating point is analyzed under Assumption \ref{assum:ss} and the corresponding exergy balance is derived from the maps in Figure \ref{fig:aval_output}. 

\begin{figure}[!t]
\centering\vspace{0.4em}
\includegraphics[width = 0.99\columnwidth]{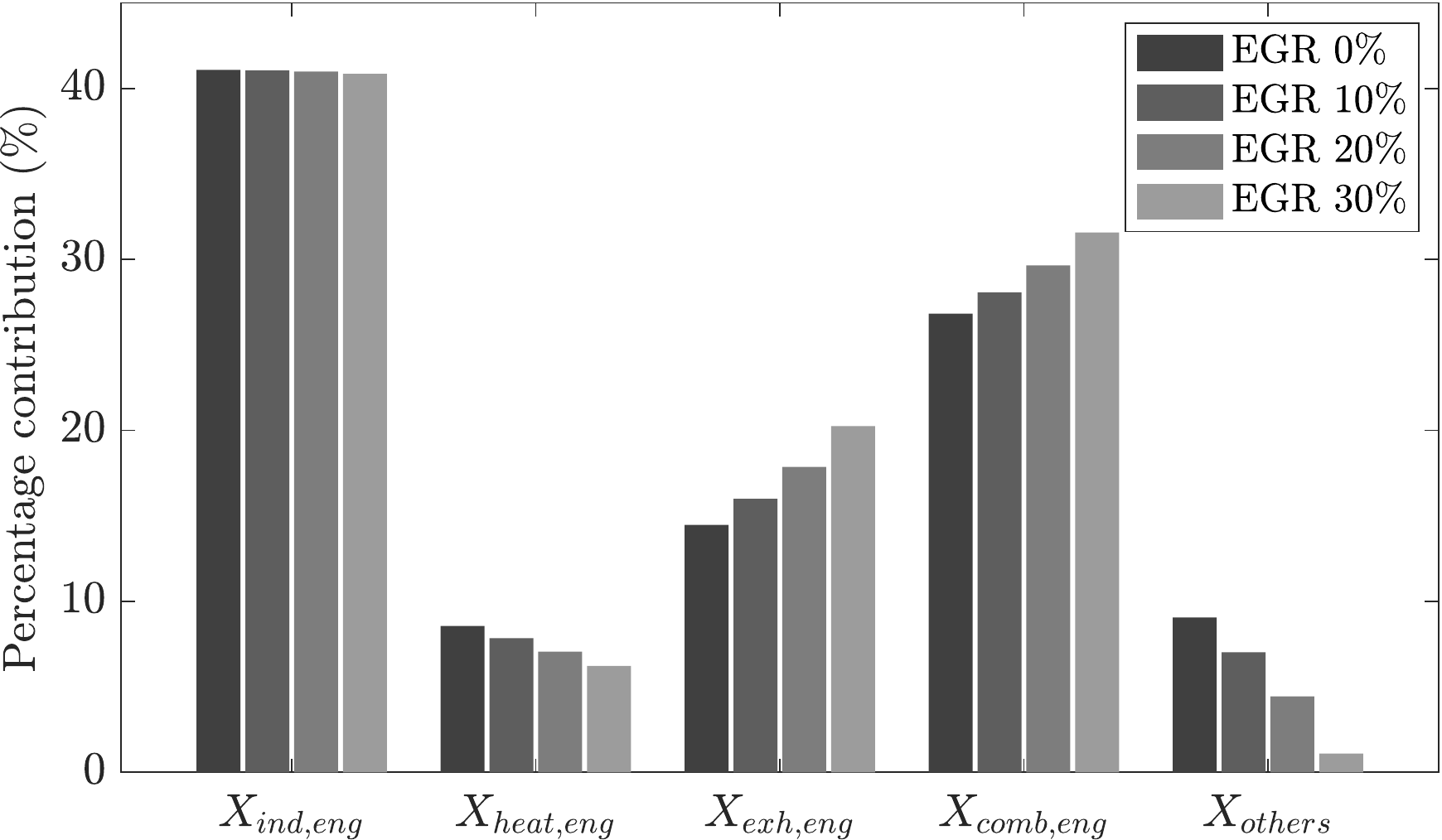}
\caption{Comparison of exergy transfer and destruction phenomena for different EGR rates.}
\label{fig:exergy_split}
\end{figure}

Given the sequence of operating point in Figure \ref{fig:pmp_hev},  Figure \ref{fig:exergy_split} shows the exergy balance simulation results considering four different EGR rates, namely: 0, 10, 20, and 30\% (a reasonable range for diesel engines \cite{rakopoulos2009diesel}).  The contribution of each exergy term is computed as:
\begin{equation}
X_k\ (\%) =  \frac{X_{k}}{X_{fuel,eng}+X_{intk,eng}}\times 100
\end{equation}
with $X_k$ obtained integrating the corresponding exergy rate term inside the set $\mathcal{K}$, introduced for Equation \myref{eq:rel_contr_ex}.  The summation of the mechanical work and friction terms ($X_{work,eng}+X_{fric,eng}$) is the indicated work $X_{ind,eng}$.

Given an engine operating point and air-fuel equivalence ratio $\lambda$, an increment of the EGR rate leads to additional recirculated exhaust gas mass and to an increase of the in-cylinder thermal inertia. This causes a decrease of the in-cylinder temperature $T_{cyl}$, of the heat transfer $\dot{Q}_{cyl}$, and, consequently, of the exergy term $X_{heat,eng}$. At lower temperatures, the combustion reaction becomes less efficient, leading to an increased entropy generation and to a higher contribution of $X_{comb,eng}$ to the total exergy balance. The additional mass of exhaust gas introduced in the cylinder must be transferred outside the system, increasing $X_{exh,eng}$. The percentage contribution of the indicated work to the overall balance is, in practice, not changing varying the EGR, with the efficiency $\varepsilon$ being 33\%. Only a slight reduction, lower than 0.5\%, is caused by the intake flux exergy increase ($X_{intk,eng}$). The term $X_{others}$ remains constrained to values in line with the literature and reaches a maximum for 0\% EGR; this latter condition is, however, unlikely for diesel engines, because it would lead to unacceptable $NO_x$ emissions\footnote{An EGR rate of 0\% can be used to model engines without exhaust recirculation. }.  

The EGR rate $x_{EGR}$, is generally a function of time and could be an additional map to be used together with the  information provided in Figure \ref{fig:map_lambda_fuel}. In this work, given that an EGR map was not available, the authors chose to use a constant EGR rate ($x_{EGR}$) and perform the sensitivity analysis above. 

\section{Conclusions}\label{sec:conclusions}
In this work, the derivation of a comprehensive mean-value exergy-based model for ICEs is provided. Starting from the mean-value modeling of the in-cylinder behavior, the exergy balance is solved for each engine operating point and static maps, describing the availability transfer and destruction as function of the engine speed and load, are obtained. For the first time, the concept of availability is extended to model the engine on all the operating points and not, as it has been done so far, a restricted set of operating points. This allows to extract useful information on the engine efficiency (and inefficiency) and quantify how moving across different operating conditions affects the exergy balance.

The method proposed in this paper has the potential of being disruptive in the vehicle control and optimization community.  The maps obtained from the application of the proposed methodology allow for a mean-value and quasi-static description of the engine's exergetic behavior in all its operating points.  Together with the framework developed in \cite{ourpaper}, this model could enable the formulation of \myquote{exergy management strategies} aiming at minimizing the overall operational losses, rather than fuel consumption only.  For example,  in the context of military vehicles, the information provided by the heat exchange exergy map could be used to develop management strategies aiming at minimizing the vehicle's thermal signature.  Exploiting the low computational burden of these static maps, the development of real-time management strategies would also be viable. 

Other than some engine technical specifications, the only information needed to compute the ICE exergy balance is the BSFC, $\lambda$, exhaust gas temperature, and EGR rate maps (typically all available to engine practitioners). If crank-angle resolved in-cylinder pressure, heat transfer, and temperature data are accessible for all the operating points, these should be used for the direct computation of maps in Figures \ref{fig:Pcyl},  \ref{fig:Tcyl}, and \ref{fig:Qdcyl}, without the need of implementing and calibrating the in-cylinder model described in Section \ref{sec:engine_dyn_crank_2}.  It is worth mentioning that,  for a given engine,  the static maps must be computed only once.

\section*{Acknowledgments}
Unclassified. DISTRIBUTION STATEMENT A. Approved for public release; distribution is unlimited. Reference herein to any specific commercial company, product, process, or service by trade name, trademark, manufacturer, or otherwise does not necessarily constitute or imply its endorsement, recommendation, or favoring by the United States Government or the Dept. of the Army (DoA). The opinions of the authors expressed herein do not necessarily state or reflect those of the United States Government or the DoD, and shall not be used for advertising or product endorsement purposes.

\appendix       
\section*{Appendix A}\label{app:proof1}
\begin{prop}\label{app_prop}
In steady-state conditions, the mean-value availability introduced in the engine at a point in time $t$, denoted by $\dot{X}_{In}$, is transferred via mass, heat, and work outside the system ($\dot{X}_{Out}$) or irreversibly destroyed ($\dot{X}_{Dest}$). The summation of these terms satisfies the following equality:
\begin{equation}
\dot{X}_{eng}(t)= \dot{X}_{In}(t) + \dot{X}_{Out}(t) + \dot{X}_{Dest}(t) = 0
\label{eq:prop}
\end{equation}
\end{prop}

\begin{remark}\label{remark_1}
Given the model developed in Section \ref{sec:ice_model} and for one engine operating point,  the quantities $\dot{X}_{In}$, $\dot{X}_{Out}$, and $\dot{X}_{Dest}$ are constant. Therefore, $\dot{X}_{eng}$ is also a constant value.
\end{remark}

\begin{remark}
Proposition A.\ref{app_prop} is formulated under the assumption that $\dot{X}_{others} = 0$. The proof proposed in this appendix can be extended to the scenario $\dot{X}_{others} \neq 0$ adding to the equality in Equation \myref{eq:prop} the term $\dot{X}_{others}$ (similarly to Equation \myref{eq:exergy_ice}). 
\end{remark}

\begin{proof} First, the proof is carried out considering one cylinder, then, results are extended to a $n_{cyl}$ cylinders engine. According to \cite{rakopoulos2009diesel}, and considering one cylinder cycle, at steady-state (i.e., at a given operating point) the following condition holds:
\begin{equation}
\int_{cycle} \frac{d\mathcal{X}_{eng}}{d\theta}d\theta = 0
\label{app_1}
\end{equation}
with \myquote{$cycle$} indicating that the integral is computed along one engine cycle, i.e., \myunit{720}{deg}. Recalling that the crank angle ($\theta$) and time ($t$) domain are linked by $d\theta = \omega_{eng} dt$, Equation \myref{app_1} is rewritten as:
\begin{equation}
\int_0^{t_{cycle}} \dot{\mathcal{X}}_{eng}(t)dt = 0
\label{app_2}
\end{equation}
with $t_{cycle}$ the time to perform one engine cycle. The mean-value model assumes that combustion takes place continuously over one cycle. Therefore, we can write the following integral:
\begin{equation}
\int_0^{t_{cycle}} \dot{X}_{eng}(t)dt = \delta
\label{app_3}
\end{equation}
with $\delta\in\mathbb{R}$. Given Remark A.\ref{remark_1}, $\dot{X}_{eng}$ is a constant value and Equation \myref{app_3} is rewritten as:
\begin{equation}
\dot{X}_{eng}\int_0^{t_{cycle}}1\ dt = \delta \rightarrow \dot{X}_{eng} = \frac{\delta}{t_{cycle}}
\label{app_4}
\end{equation}
From Equation \myref{app_2}, we know that the exergy balance over one cycle is zero, consequently:
\begin{equation}
\dot{X}_{eng} = \frac{\delta}{t_{cycle}}=0
\label{app_5}
\end{equation}
Considering $n_{cyl}$ cylinders working at steady-state, and recalling Assumption \ref{assum:ss_incyl}, Equation \myref{app_5} can be easily extended as follows:
\begin{equation}
\dot{X}_{eng} = n_{cyl}\times\frac{\delta}{t_{cycle}}=0
\end{equation}
\hfill$\square$
\end{proof}

The previous proof is developed for one engine cycle. Considering $N$ cycles, the following result holds: $\dot{X}_{eng} = N\times\left(n_{cyl}\delta/t_{cycle}\right)=0$.

\section*{Appendix B}\label{app:proof2}
\begin{figure}[H]
\centering
\includegraphics[width = \columnwidth]{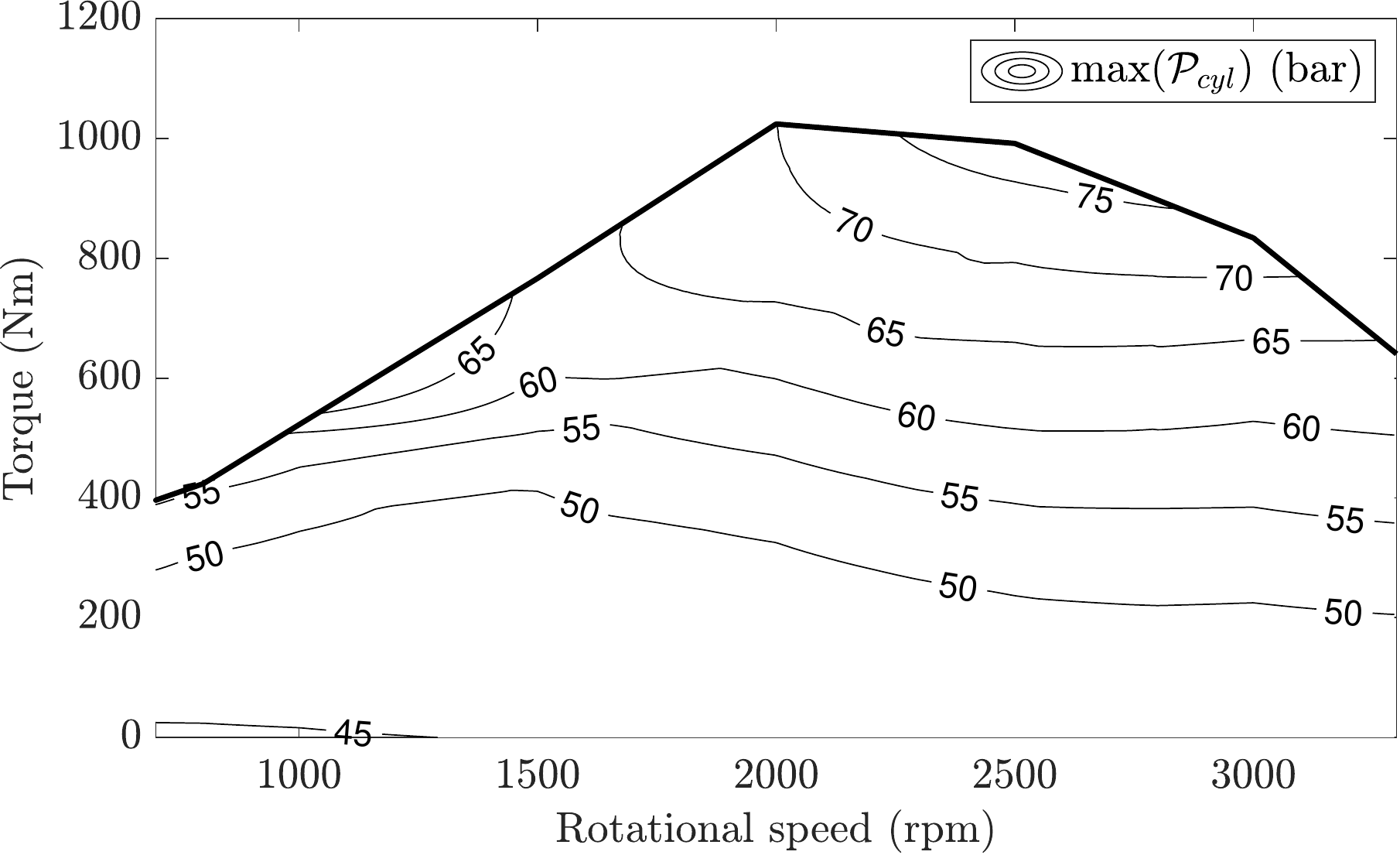}
\caption{Maximum of $\mathcal{P}_{cyl}$ for each engine operating point. Results are obtained considering the parameters listed in Table 1.}
\label{fig:supp_1}
\end{figure}
\vspace{-1em}

\begin{figure}[H]
\centering
\includegraphics[width = \columnwidth]{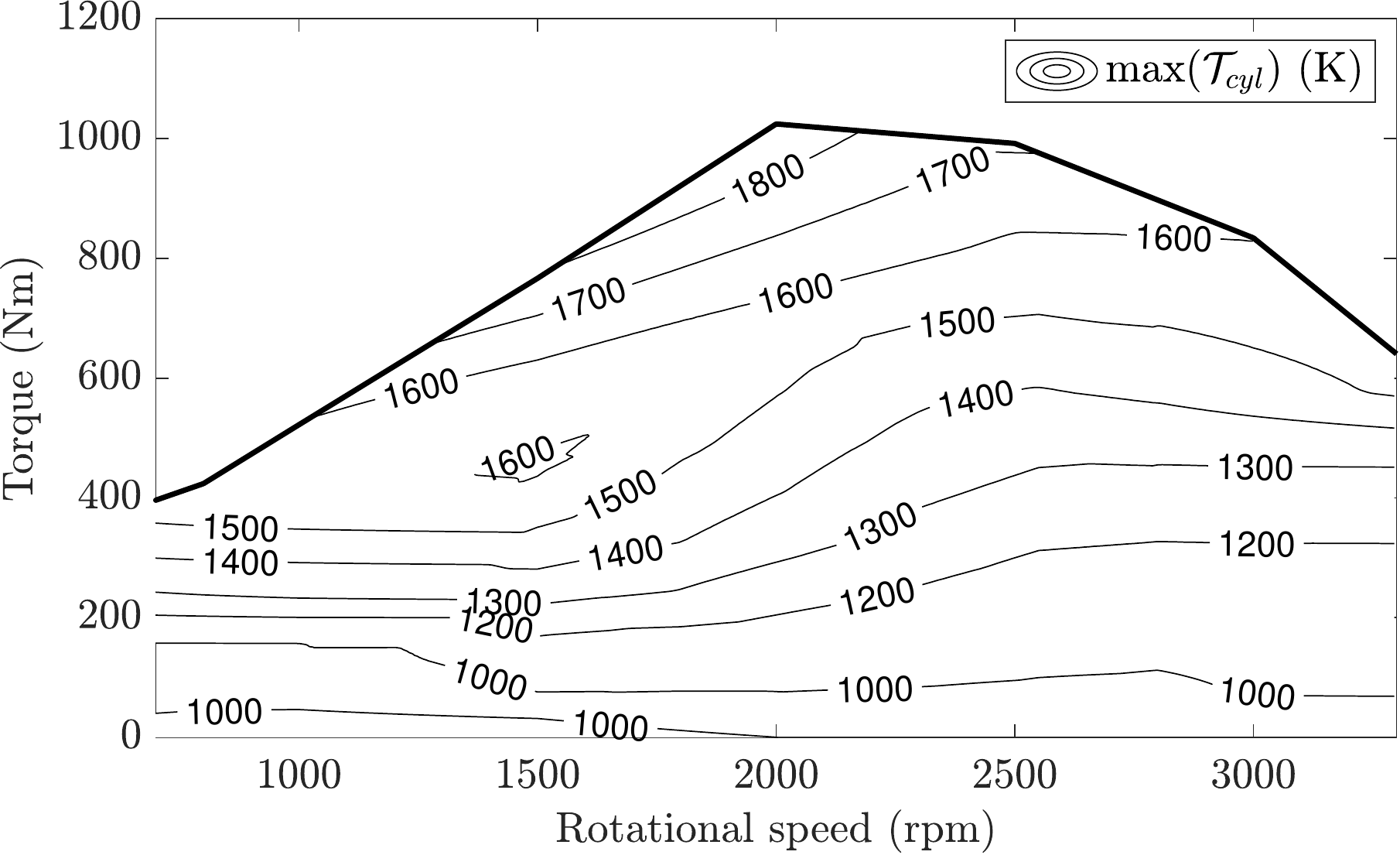}
\caption{Maximum of $\mathcal{T}_{cyl}$ for each engine operating point. Results are obtained considering the parameters listed in Table 1.}
\label{fig:supp_2}
\end{figure}
\vspace{-1em}

\begin{figure}[H]
\centering
\includegraphics[width = \columnwidth]{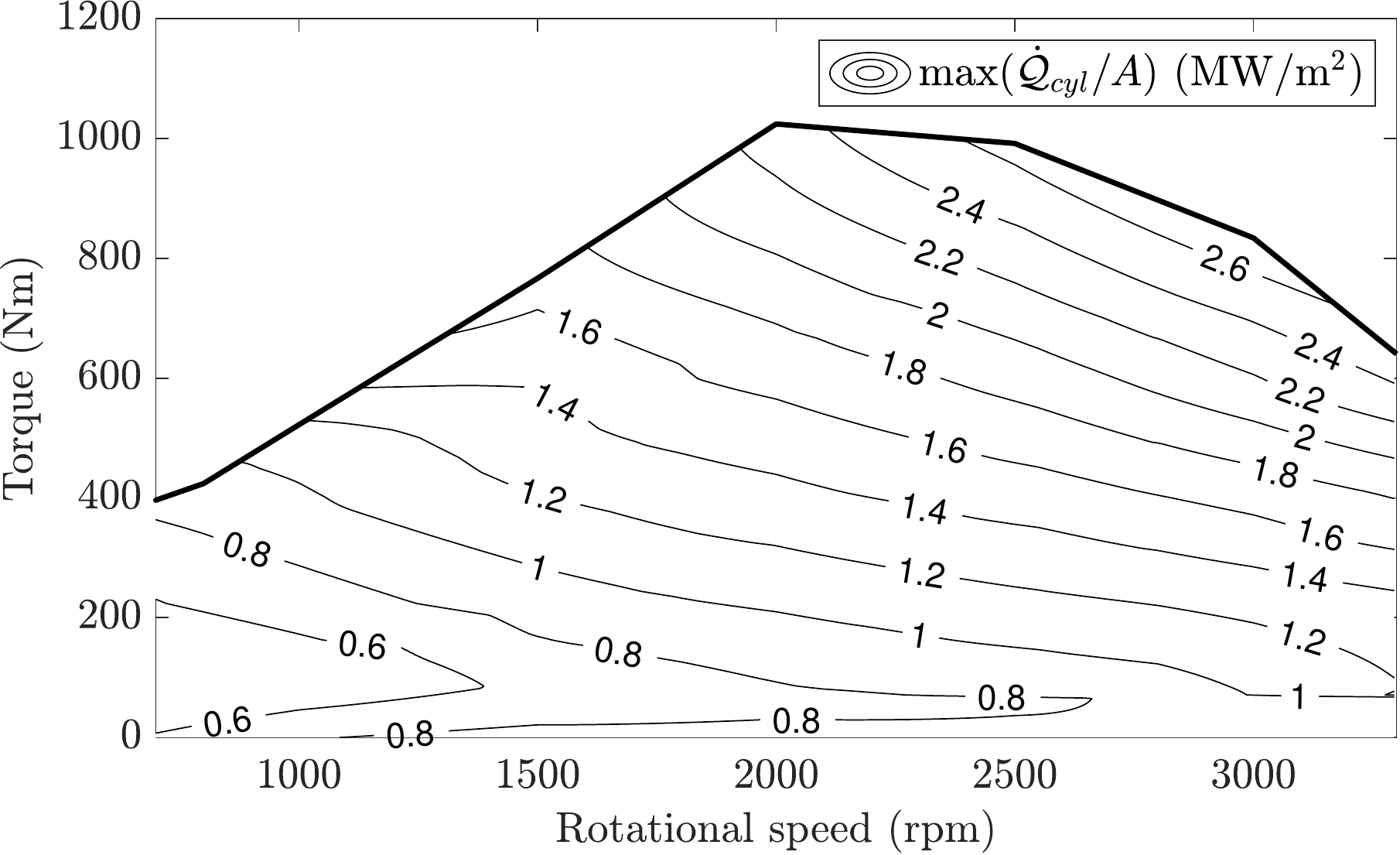}
\caption{Maximum of $\dot{\mathcal{Q}}_{cyl}/A$ for each engine operating point. Results are obtained considering the parameters listed in Table 1.}
\label{fig:supp_3}
\end{figure}

\bibliographystyle{asmems4}
\bibliography{biblio}

\end{document}